\documentclass[preprint]{iucr}              % DO NOT DELETE THIS LINE

     \journalcode{A}              % Indicate the journal to which submitted
\usepackage{graphicx}
\usepackage{xcolor}

\begin{document}                  % DO NOT DELETE THIS LINE

     %-------------------------------------------------------------------------
     % The introductory (header) part of the paper
     %-------------------------------------------------------------------------

     % The title of the paper. Use \shorttitle to indicate an abbreviated title
     % for use in running heads (you will need to uncomment it).

\title{A Group-theoretical Approach to Enumerating Magnetoelectric and Multiferroic Couplings in Perovskites}
%\shorttitle{Short Title}

     % Authors' names and addresses. Use \cauthor for the main (contact) author.
     % Use \author for all other authors. Use \aff for authors' affiliations.
     % Use lower-case letters in square brackets to link authors to their
     % affiliations; if there is only one affiliation address, remove the [a].
		
\cauthor[a]{Mark S.}{Senn}{m.senn@warwick.ac.uk}{address if different from \aff}
\author[b,c]{Nicholas C.}{Bristowe}

\aff[a]{Department of Chemistry, University of Warwick, Gibbet Hill, Coventry, CV4 7AL \country{United Kingdom}}
\aff[b]{School of Physical Sciences, University of Kent, Canterbury CT2 7NH, \country{United Kingdom}}
\aff[c]{Department of Materials, Imperial College London, London SW7 2AZ, \country{United Kingdom}}

%\keyword{keyword}

     % PDB and NDB reference codes for structures referenced in the article and
     % deposited with the Protein Data Bank and Nucleic Acids Database (Acta
     % Crystallographica Section D). Repeat for each separate structure e.g
     % \PDBref[dethiobiotin synthetase]{1byi} \NDBref[d(G$_4$CGC$_4$)]{ad0002}

%\PDBref[optional name]{refcode}
%\NDBref[optional name]{refcode}

\maketitle                        % DO NOT DELETE THIS LINE

\begin{synopsis}
A symmetry motivated approach for designing perovskites with ferroic and magnetoelectric couplings is proposed. The results highlight which kinds of magnetic orderings and structural distortions need to coexist within the same structure to produce the desired couplings.
\end{synopsis}

\begin{abstract}

We use a group theoretical approach to enumerate the possible couplings between magnetism and ferroelectric polarisation in the parent $Pm\bar{3}m$ perovskite structure.  We show that third order magnetoelectric coupling terms must always involve magnetic ordering at the A and B-site which either transforms both as R-point or both as X-point time odd irreducible representations (irreps).  For fourth order couplings we demonstrate that this criterion may be relaxed allowing couplings involving irreps at X, M and R-points which collectively conserve crystal momentum, producing a magnetoelectric effect arising from only B-site magnetic order.  In this case, exactly two of the three irreps entering the order parameter must be time-odd irreps and either one or all must be even with respect to inversion symmetry. We are able to show that the time-even irreps in this triad must transform as one of: X$_{1}^{-}$, M$_{3,5}^{-}$ and R$_{5}^{+}$, corresponding to A-site cation order, A-site anti-polar displacements or anion rock-salt ordering.  This greatly reduces the search-space for (type-II multiferroic) perovskites. We use similar arguments to demonstrate how weak ferromagnetism may be engineered, and we propose a variety of schemes for coupling this to ferroelectric polarisation.  We illustrate our approach with DFT calculations on magnetoelectric couplings, and by considering the literature we suggest which avenues of research are likely to be most promising in the design of novel magnetoelectric materials.

\end{abstract}

\section{Introduction}

The classification of distortions in functional materials is an important part of the process of understanding the structure-property relationship.  Perovskites (ABX$_3$) are among the most studied systems, which is in part due to the many functional properties that they exhibit, but also due to their richness in structural distortions and phase transitions. Schemes classifying the ubiquitous rotations and tilts of the quasi-rigid BO$_6$ octahedra that drive many of these phase transitions in perovskites, can be conveniently classified in terms of Glazer notation~\cite{Glazer1972}, and other such schemes also exits for classifying distortions in layered perovskite such at Ruddlesden-Poppers~\cite{Aleksandrov1994}.  While these schemes have enjoyed much success due to their intuitive nature, there are several limitations, in particular that they are not easily generalised to different systems. Even within the perovskite family with additional symmetry breaking with respect to the ABX$_3$ aristotype, it is no longer clear how the occurrence of tilts and rotations can be unambiguously described, or indeed how the symmetry lowering implied by the combined orderings can be derived.    

More formally, the degrees of freedom in an aristotype ``parent'' structure, such as the $Pm\bar{3}m$ $ABX_3$ perovskite, may be defined as transforming as irreducible representations (irreps) of the parent space group (and setting).  The irreps for all special positions in reciprocal space have been tabulated by various authors including by Bradley and Cracknell \cite{Bradley1972}, Miller and Love \cite{Miller1967}, Kovalev \cite{Kovalev1993} and more recently also at non-special k-points~\cite{Stokes2013}.  With knowledge of these irreps, it is possible to compute the isotropy subgroups of the 230 space groups,~\cite{Stokes1988} which are the subgroups accessible due to the action of an order parameter transforming as one of these irreps.

Online tools such as $ISODISTORT$\cite{Campbell2006} and $AMPLIMODES$ on the Bilbao Crystallographic server~\cite{Aroyo2006,Orobengoa2009} allow distorted structures to be easily decomposed in terms of irreps of a parent space group, and it is now possible to superpose up to 3 irreps with associated independent incommensurate propagation vectors, and derive the possible subgroups and secondary order parameters \cite{Stokes2017}.  Additionally, these programs now generate outputs that can be directly read by Rietveld and single crystal refinement programs \cite{Campbell2007,PerezMato2010} allowing refinements to be performed in a symmetry adapted basis and facilitating easy identification of the active order parameters in a given phase transition.

As a result of much of this work, several group-theoretical studies have emerged that have more formally classified distortions in Perovskite related materials. These include group-theoretical analysis of octahedral tilting in perovskites \cite{Howard1998,Howard2005,Knight2009}, cation ordered and Jahn-Teller distortions in perovskites \cite{Howard2010}, ferroelectric perovskites \cite{Stokes2002}, anion ordering \cite{Talanov2016}, and works on layered Ruddlesden-Poppers \cite{Hatch1987,Hatch1989}.  One particularly valuable aspect of classifying these distortions in the formal language of irreps is to understand physical phenomena that can arise due to secondary order parameters which feature at linear order in the Landau-style free energy potential. These odd order terms may always adopt a sign such that they act to lower the overall free energy and hence symmetry analysis alone is sufficient to identify their instability. The process of ascertaining these couplings is greatly simplified using the ideas of invariants analysis  \cite{Stokes1991,Saxena1994} when constructing the Landau-style free energy expansion about the parent undistorted phase, and online tools for doing this also exist \cite{Hatch2003}. 

This process is particularly valuable when understanding improper ferroelectricity ~\cite{Levanyuk1974} where third order terms in the free energy expansion are invariably the key to understanding the resulting polarisation.  This area has enjoyed a renascence in the form of the recently much discussed ''hybrid improper ferroelectric'' mechanism (e.g. see ~\cite{Benedek2015} for a recent review).  The powerful use of magnetic superspace groups for describing multiferroic materials has also allowed magnetoelectric couplings to be trivially identified through analysis of secondary order parameters \cite{Perez-Mato2012}. Antisymmetric exchange arguments with respect to the parent perovskite structure have also been used to explain the dominant anisotropic terms that control the directions of spin ordering \cite{Khalyavin2015a}.  And of course, the occurrence of weak ferromagnetism (wFM) by the Dzyaloshinsky-Moriya interaction \cite{Dzyaloshinsky1958,Moriya1960} was first originally rationalised based on such symmetry arguments alone \cite{Dzyaloshinsky1958}.

Using many of these ideas above, and with the aid of the $ISODISTORT$\cite{Campbell2006} tool, we seek here to generalise a recipe for inducing magnetoelectricity in the parent $Pm\bar{3}m$ perovskite.  These recipes are based on symmetry arguments alone, and we use as the ingredients structural and magnetic degrees of freedom, which we classify in terms of transforming as irreducible representations of the parent space group. Our results clearly show why certain kinds of coupled distortions and magnetic ordering can never lead to ferroelectric or ferromagnetic secondary order parameters, and by considering which orderings and cation arrangements are commonly observed, we are able to identify several promising avenues for further investigation.   

The manuscript is arranged as follows. In section 2, we first classify the ingredients for symmetry breaking that are at our disposal in terms of irreps of the parent $Pm\bar{3}m$ space group.  To keep our results as general as possible, we will also describe cation and anion ordering in terms of irreps, rather than forming new parent space groups.   We then proceed to give various recipes for achieving (multi)ferroic orderings as a consequence of different symmetry breaking distortions. In Section 3, for completeness we give the recipe for (hybrid) improper ferroelectricity, while in Sections 4,5,6 we discuss magnetoelectric couplings arising due to third and fourth order terms in the free energy expansion. As the most useful multiferroics are those that are ferromagnets (rather than antiferromagnets), in Section 7 we explain how similar ideas can be used to design systems that exhibit weak ferromagnetism (wFM).  We also consider in this section systems in which either P or wFM is supplied as an external order parameter (as a magnetic or electric field) resulting in the development of wFM or P respectively in response to the stimuli.  Finally in Section 8 we put all of our above ideas together and deal with the design of materials that are both wFM and ferroelectric, and have indirect coupling through at least one primary order parameter.

\section{Ingredients for symmetry breaking}

\begin{figure}
\caption{Basic AFM magnetic orderings of the Perovskite structure with associated irrep labels, and illustrated along high symmetry order parameter directions. A-sites, B-sites and X-sites are shown as green, red and blue spheres respectively.  The parent cubic unit cell is shown in pink so as to illustrate the relationship with the new crystallographic axes (gray).  All figures are drawn in $ISODISTORT$.}
\includegraphics{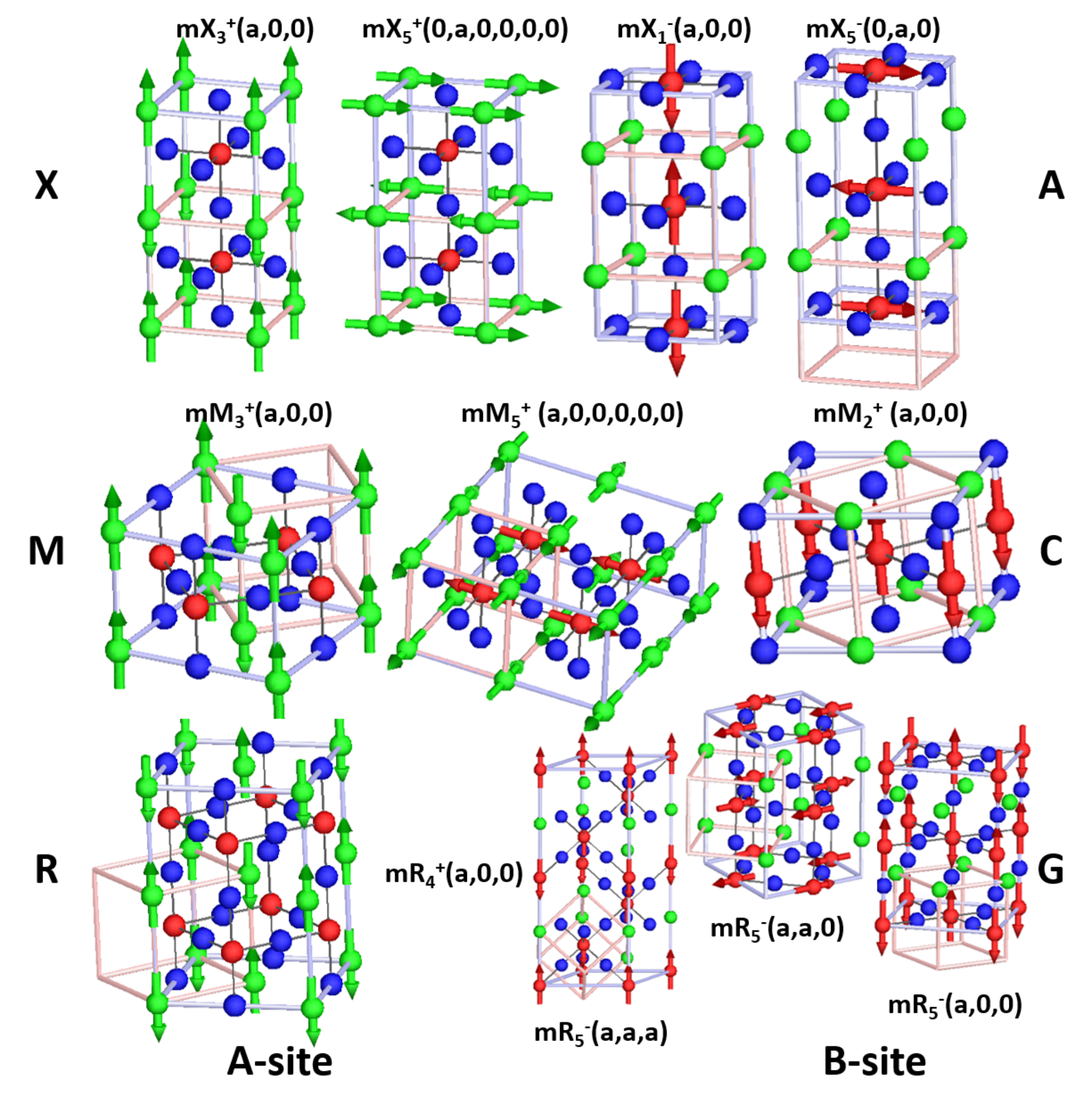}
\label{magneticmodes}
\end{figure}

First we classify the magnetic degrees of freedom at our disposal in terms of irreps of the space group $Pm\bar{3}m$.  We classify all of these in terms of irreducible representations of the parent perovskite structure $Pm\bar{3}m$ with setting  A 1a (0,0,0); B 1b ($\frac{1}{2}$,$\frac{1}{2}$,$\frac{1}{2}$);  X 3c (0,$\frac{1}{2}$,$\frac{1}{2}$). We note that reversing the setting of the structure will result in many of the irrep labels changing, in particular at the X and R points, irreps labelled as ``+'' will correspond to another numbered irrep with ``-'' sign and visa versa.  The origin of this is that the sign part in these irrep labels refer to whether or not parity (with respect to inversion symmetry) is conserved or violated at the origin (0,0,0), and hence interchanging the atom at the origin naturally effects the distortions physically being described by a particular representation. The orderings of the magnetic degrees of freedom will ultimately be devised in such a way as to drive secondary order parameters that are related to ferroelectricity.  We restrict ourselves here to the basic types of antiferromagnetic ordering which are commonly observed in perovskites.  These are often characterised as A, C and G-type having 1, 2 and 3 AFM nodes respectively.  They may be classified as corresponding to orderings which transform as irreps at the X[0, $\frac{1}{2}$, 0], M[$\frac{1}{2}$, $\frac{1}{2}$, 0] and R[$\frac{1}{2}$, $\frac{1}{2}$, $\frac{1}{2}$] point (Figure~\ref{magneticmodes}).  It is important to note that magnetic structures such as A$_{x}$ and A$_{yz}$, which correspond to an ordering with propagation vector X[$\frac{1}{2}$,0,0] with moment along the propagation axis and perpendicular to it, transform as distinct irreps in this analysis, and will imply physically distinct secondary order parameters. This forms the basis of the antisymmetric exchange arguments of Khalyavin \cite{Khalyavin2015a} to determine spin (exchange) anisotropy, and this is why this analysis is so powerful in the perovskite structure where the magnetic atoms sit on high symmetry sites.  Figure~\ref{magneticmodes} gives full details of how the spin arrangements are related to irreps.

Next we classify the various structural degrees of freedom within the perovskite structure for inducing symmetry lowering phase transitions.  The ingredients at our disposal are the commonly observed octahedral rotation and tilt modes, Jahn-Teller distortion modes, cation (charge) ordering modes, anti-polar modes and strain. These are all listed in Table~\ref{ingredients}, along with their corresponding labels in the alternative setting (A at ($\frac{1}{2}$,$\frac{1}{2}$,$\frac{1}{2}$)).  Some of these degrees of freedom will be accessible via physical control parameters (such as application of epitaxial strain) whilst others only by chemical design (for example by inclusion of Jahn-Teller active cations).  In the above analysis, we will also classify cation and anion orderings in the perovskite structure in terms of transforming as irreducible representations of the parent perovskite.  For example, rock salt cation ordering at the B-site transforms as R$_{2}^{-}$ and A-site layered cation order as X$_{1}^{+}$.  We may even classify the highly distorted cation ordered A'A$_3$B$_4$O$_{12}$ quadruple perovskite with aristotype $Im\bar{3}$, as having cation orderings transform as  M$_{1}^{+}$ (with three k-actives = [$\frac{1}{2}$, $\frac{1}{2}$, 0];[0,$\frac{1}{2}$, $\frac{1}{2}$];[$\frac{1}{2}$, 0, $\frac{1}{2}$]) and octahedral rotations that stabilise the A' square planar coordination transforming as M$_{2}^{+}$.

Finally, the desired property, ferroelectricity, transforms as the polar mode belonging to the irrep $\Gamma_{4}^{-}$.  $\Gamma_{4}^{-}$ is a three dimensional irrep, the most general order parameter direction (OPD) associated with this would hence be written as OP(a,b,c), where special directions, (a, 0, 0), (a, a, 0) and (a, a, a,) correspond to tetragonal, orthorhombic and rhombohedral directions for the macroscopic polarisation and off-centre displacements of the atoms. For a full discussion on notation relating to OPD, including cases where multiple irreps enter in to the OP, as will become pertinent in future discussion, the reader is directed to Appendix A. Please note that throughout this manuscript we choose to list the full OPD, instead of the space group and setting. The two are equivalent, but we choose the OPD for the sake of brevity, and also due to its descriptive nature with respect the magnetic and structural orderings that are all allowed to occur. We will now discuss the general design principles by which we can combine the aforementioned degrees of freedom to produce $\Gamma_{4}^{-}$ as a secondary order parameter.

\section{Recipes for Improper Ferroelectric Couplings}

We begin by considering structural irreps (transforming as time even) alone, and how they may combine to produce improper ferroelectric couplings, before considering couplings with magnetic irreps in the next section. The concept of improper ferroelectricity was first introduced several decades ago by Levanyuk~\cite{Levanyuk1974}, but recently there has been renewed interest (see reviews \cite{Varignon2015a,Benedek2015,Young2015a}) after its observation in epitaxially grown layered perovskite systems~\cite{Bousquet2008}. In light of work that has highlighted the existence of improper ferroelectricity in naturally layered perovskite-like Ruddlesden-Popper systems~\cite{Benedek2011}, we believe it is also of interest to enumerate all such possible couplings in the aristotypical perovskite structure here, at least to illustrate the idea, introduce the topic and review the literature, before moving on to magnetoelectric couplings. 

The general recipe for constructing improper ferroelectric coupling terms in the Landau-style free energy expansion about the parent perovskite structure that we will use is as follows.  The principle of Invariants analysis~\cite{Hatch2003} means that, at each term in the free energy expansion, crystal momentum and inversion symmetry must be conserved. In the next section we also consider magnetism, when the additional constraint of time reversal symmetry must be conserved.  

We seek initially the dominant coupling term, which means that we should consider the lowest order term in the free energy expansion that is achievable which has linear order in polarisation (P). We restrict ourselves to coupling terms only of linear order in P since in these cases we can be sure that symmetry analysis can be sufficient to infer the appearance of P, unlike in even orders where calculation of the sign and strength of the coefficients would be necessary. For example, since P transforms as inversion odd and has zero crystal momentum, the lowest order term will be third order (ABP), which has been termed hybrid improper ferroelectricity~\cite{Bousquet2008,Benedek2011,Fukushima2011b}. Since trilinear terms will always act to lower the free energy, if A and B are unstable, then P will also be present, adopting a sign (direction of polarisation) such as to stabilise the overall free energy.

Invariants analysis tells us that: 

for P is inversion-odd; [P] = [0,0,0]~\footnote{In general the inversion symmetry breaking distortion will transform as the polar $\Gamma_{4}^{-}$ mode, however sometimes the symmetry breaking will instead be associated with another $\Gamma^{-}$ mode which is piezoelectric in nature}.

A$\cdot$B is inversion-odd; [A]+[B] = [0,0,0]  must be obeyed leading to all quantities being conserved in the trilinear term:

A$\cdot$B$\cdot$P is inversion-even; [A]+[B]+[P] =[0,0,0] to be true, where [A] represents crystal momentum associated with order parameter A and A$\cdot$B is the dot product of the irreps associated with the OP A and B.~\footnote{Strictly speaking, the relevant OPs are vectors whose elements (real or imaginary numbers) that reflect the amplitude of the atomic displacements or magnetic moments that transform according to specific irreps (which are themselves vectors). However, since for the purposes of our symmetry analysis, the information we require concerning crystal momentum and parity is encoded in the irrep label (and we are not concerned with amplitude here), we will also label the OPs using this notation}

One may further convince oneself that A $\neq$ B must be true for this condition to be fulfilled for otherwise AB would be inversion-even, meaning that the quadratic-linear term A$^{2}$P is not permissible in the free energy expansion, and so is not a term that can drive an improper coupling~\footnote{A$^{2}$P terms can be possible in some systems where A transforms as an irrep with imaginary character, but this is not relevant for the zone-boundary irreps of $Pm\bar{3}m$ that we consider here.}. In summary we can say that A and B must both be of opposite parity with respect to inversion symmetry and must have equal crystal momentum. We will explore all trilinear couplings possible within the perovskite parent structure for order parameters transforming as X, M and R-point irreps below. 

The above criteria is necessary, but in a few cases not always sufficient to ensure the desired improper ferroelectric coupling. In practice this may be conveniently checked using ``Method 2'' of the online tool ISODISTORT where multiple irreps. may be superimposed to form the primary order parameter of the parent perovskite structure.  The program then lists all the possible order parameter directions associated with this, along with the resulting secondary order parameters and the space group symmetry and basis with respect to the parent structure. It is then trivial to identify from either the space group or the list of secondary order parameters if an improper ferroelectric coupling will occur.

Any of the following that have atomic displacements that transform collectively as these irreps, will feature in a trilinear term with $\Gamma_{4}^{-}$ (where $\oplus$ represents the direct sum):

X$_{1,2,3}^{+}$ $\oplus$ X$_{3,5}^{-}$ \\

M$_{1,2,3,4,5}^{+}$ $\oplus$ M$_{2,3,5}^{-}$ \\

R$_{1,5}^{+}$ $\oplus$ R$_{2,3,4,5}^{-}$ \\

While many of these may be difficult to achieve in practice, there are several promising candidates.  For example, layered A-site cation order (M$_{1}^{+}$) with anti-polar B-site displacements (M$_{5}^{-}$) can lead to a trilinear term M$_{1}^{+}$ M$_{5}^{-}$ $\Gamma_{4}^{-}$.  We believe this could be the cause of the ferroelectric polarisation recently reported in high pressure perovskite CaMnTi$_2$O$_6$~\cite{Aimi2014}. Indeed, cation or anion ordering at any of the perovskite sites at the M-point along with anti-polar distortions at the A or B-sites would produce an improper ferroelectric polarisation. In-phase tilting (M$_{2}^{+}$) or the M-point Jahn-Teller mode (M$_{3}^{+}$) can alternatively be used in conjunction with the anti-polar displacements (such as M$_{5}^{-}$) to induce a polarisation, which has been recently predicted in the $Pmc2_1$ phase of several perovskites~\cite{Yang2012,Yang2014,Varignon2016}, and might also be the origin of the (ionic component of the) ferroelectricity in the $P2_1nm$ half-doped manganites~\cite{Giovannetti2009,Rodriguez2005}. 

The commonly observed rock-salt cation ordering at the B-site \cite{King2010} along with (R-point) anti-polar distortions on the A-site will also produce an improper ferroelectric coupling.  While the former is commonly observed, controlling the periodicity of the anti-polar distortions such as those induced by lone pair ordering will be challenging. Cation order on the A-sites at the R-point (rock-salt) along with octahedral tilt modes would also produce an improper ferroelectric coupling, as recently predicted through first principles calculations~\cite{Young2013}. However, it should be noted that A-site cation ordering is more commonly found to be in a layered (X-point) arrangement \cite{King2010}.  Very recent reports of improper ferroelectricity in the 134-perovskites HgMn$_{3}$Mn$_4$O$_{12}$ can also be understood with respects to the present symmetry analysis of AB$O_3$ perovskites~\cite{Chen2018}.  In this case, the atomic displacements associated with the orbital and charge ordering degrees of freedom on the A and B-sites transform as irreps of the parent space group $Pm\bar{3}m$ R$_{5}^{+}$ and  R$_{3}^{-}$.

A-site cation layering (X$_{1}^{+}$) in combination with anti-polar A-cation motions is indeed sufficient to induce P. Again, whilst the former is fairly common, the latter is only expected to be an unstable lattice distortion for low tolerance factor perovskites \cite{Mulder2013}. However it can itself manifest through an improper appearance with two tilting modes (M$_{2}^{+}$ R$_{5}^{-}$ X$_{5}^{-}$), which gives rise to the fourth order term described below. At the X-point, one other trilinear term has been predicted to play a role in the $P2_{1}$ phase of strained CaTiO$_{3}$, whereby A and B-site antipolar (X$_{5}^{+}$ and X$_{5}^{-}$) motions induce P~\cite{Zhou2013}.

Fourth order terms in P should also be considered and may be more promising on account of the extra degree of flexibility allowed in the recipe~\footnote{Although we consider these formally as fourth order terms here we emphasise that they might equally be third order terms of a lower symmetry perovskite structure that has already undergone some cation ordering or other structural distortion}. Here, crystal momentum considerations mean that each relevant fourth order term must take the form:

A$\cdot$B$\cdot$C is inversion-odd; [A]+[B]+[C] = [0,0,0]  must be obeyed leading to all quantities being conserved in the trilinear term:

A$\cdot$B$\cdot$C$\cdot$P is inversion-even; [A]+[B]+[C]+ [P] = [0,0,0] to be true 

One of the most promising fourth order candidates involves order parameters associated with X$^{+}$, M$^{+}$, R$^{-}$ and $\Gamma_{4}^{-}$.  For example, A-site layered cation ordering (X$_{1}^{+}$), octahedral tilt mode (M$_{2}^{+}$) and octahedral tilt mode (R$_{5}^{-}$). This explains the significance of layering (X$_{1}^{+}$) in allowing the two octahedral rotation modes to couple together to produce a polarisation and has been the most common example of improper ferroelectricity in perovskites as illustrated in both artificially~\cite{Bousquet2008,Rondinelli2012}, and naturally layered double perovskites \cite{Fukushima2011b}. A similar term, predicted in half-doped titanates~\cite{Bristowe2015}, includes A-site layered cation ordering (X$_{1}^{+}$), M-point Jahn-Teller (M$_{3}^{+}$) and octahedral tilt modes (R$_{5}^{-}$). Other possibilities include A-site striped cation ordering (X$_{1}^{+}$), tilting (R$_{5}^{-}$) and charge order (M$_{4}^{+}$), which we believe to be the origin of the improper polarisation in SmBaMn$_2$O$_6$~\cite{Yamauchi2013}. Alternatively Jahn-Teller induced, M$_{3}^{+}$ and R$_{3}^{-}$, ferroelectricity has been discussed in A-site striped cation ordered (X$_{1}^{+}$) rare-earth vanadates~\cite{Varignon2015}. Perhaps an interesting avenue for future research is to use anion ordering since the X$_{1}^{+}$ irrep is also made possible by anion vacancy ordering, which for example is sometimes seen in the cobaltates~\cite{Karen2001,Vogt2000,Castillo-Martinez2006}. 

Other chemically and structurally less promising schemes are still worth a mention. X$^{-}$ M$^{-}$ R$^{-}$ $\Gamma_{4}^{-}$, for example, striped order at A-site (X), anti-polar order at B-site (M), and rock-salt cation order at B-site (R); X$^{-}$ M$^{+}$ R$^{+}$ $\Gamma_{4}^{-}$, striped B-site cation order (X$_{3}^{-}$), Octahedral tilt mode (M$_{2}^{+}$), anti-polar distortion on the B-site (R$_{5}^{+}$) and X$^{+}$ M$^{-}$ R$^{+}$ $\Gamma_{4}^{-}$ , A-site striped cation ordering (X$_{1}^{+}$), Anti-polar distortions on the B-site (M$_{2}^{-}$), anion order (R$_{5}^{+}$).  Finally, we note that the inclusion of organic cations on the A-site or organic link molecules on the X-site greatly increases the possible number of such improper ferroelectric coupling scheme, \cite{Bostrom2017} and provides a promising route for designing novel functional materials.

\section{Recipes for Magnetoelectric coupling}

We can extend the ideas discussed above for improper ferroelectrics, to magnetoelectric couplings including time-odd irreps that describe magnetic order. We seek initially the strongest magnetoelectric coupling term possible, this means that as before we should consider the lowest order term in the free energy expansion that is achievable.  Since P transforms as time even, inversion odd and has zero crystal momentum, the lowest order term involving two zone-boundary irreps will be third order (ABP).  Invariants analysis tells us that: 
for P is time-even; P is inversion-odd; [P] = [0,0,0]

A$\cdot$B is time-even; A$\cdot$B is inversion-odd; [A]+[B] = [0,0,0]  must be obeyed leading to all quantities being conserved in the trilinear term:

A$\cdot$B$\cdot$P is time-even; A$\cdot$B$\cdot$P is inversion-even; [A]+[B]+[P] =[0,0,0] to be true 

As we are seeking a magnetoelectric coupling, at least one of A or B must be magnetic, and inspection of the condition that A$\cdot$B is time-even, means that therefore both A and B must transform as a time-odd irreducible representation.  One may further convince oneself that A $\neq$ B must be true for this condition to be fulfilled for otherwise AB would be inversion-even, meaning that the quadratic-linear term A$^{2}$P is not permissible in the free energy expansion, and so is not a term that can drive an electromagnetic coupling~\footnote{A$^{2}$P terms can be possible in some systems where A transforms as an irrep with imaginary character, but this is not relevant for the zone-boundary irreps of $Pm\bar{3}m$ that we consider here.}. Taking everything together we can say that A and B must both be time-odd, of opposite parity with respect to inversion symmetry and must have equal crystal momentum. 

As before with the improper ferroelectrics, the list of magnetoelectric trilinear coupling terms (with respect to the perovskite parent structure) will prove to be rather restrictive, and so we will also consider fourth order terms in the free energy expansion.  This would be equivalent to considering trilinear terms of a new parent structure which has one of the many reported subgroups of $Pm\bar{3}m$ due to structural distortions or cation orderings, which themselves can be classified as transforming as irreps of $Pm\bar{3}m$.  However, from a materials design perspective, it is most convenient to always list these couplings with respect to the aristotypical symmetry.

If we consider couplings at the fourth order we may now construct terms as follows from the three primary order parameters (A, B and C):

A$\cdot$B$\cdot$C is time--even 

A$\cdot$B$\cdot$C is inversion--odd

[A] + [B] + [ C] = [0, 0, 0]

If we are seeking a magnetoelectric coupling, precisely two of these terms must be time-odd (since P will always be time-even), but the constraint that the sum of these two terms must conserve crystal momentum is now lifted. We will refer to this design approach as ``closing the momentum triangle'', since now three vectors (irreps) may be chosen to produce zero crystal momentum transfer.

This gives greater flexibility in the design strategy, but the price of course is that now three primary OP are required.  This means either these must all spontaneously become thermodynamically favourable at the phase transition, or more likely, and as discussed above, the structure will already contain distortions to the parent phase (such as octahedral rotations) which are ubiquitous in the perovskite structure.   

Our approach outlined above is similar in spirit in some manners to that used to consider possible magnetoelectric couplings in the incommensurate phase of BaMnF$_4$ ~\cite{Fox1980}.  However, our approach differs in that we perform the Landau-style expansion of the free energy about a hypothetical aristotypical symmetry, rather than the experimentally observed high temperature phases.   The benefit of our approach is that it encodes as much information as possible regarding the crystal momentum and parity of the time odd and even order parameters into the problem, making it particularly easy to predict magnetoelectric couplings based on symmetry arguments alone, as we demonstrate here.

\section{Trilinear Magnetoelectric couplings in AFM systems}

We start from the criteria derived above which mean that we may superpose the following time-odd irreps when constructing the order parameter:

mR$_{4}^{+}$ $\oplus$ mR$_{5}^{-}$; mX$_{3}^{+}$ $\oplus$ mX$_{1}^{-}$; mX$_{5}^{+}$ $\oplus$ mX$_{5}^{-}$

At the M-point, all possible magnetic orderings transform as mM$^{+}$ and so no magnetoelectric couplings are possible. This finding immediately rules out a large area of search space. Furthermore, magnetic moments on the A-site cations transform always as mX$^{+}$ and mR$^{+}$ and B-sites always as mX$^{-}$ and mR$^{-}$, meaning any such trilinear magnetoelectric coupling mechanism must involve order on both A and B-sites simultaneously.  We take these three possible couplings in turn now, and consider which are the most physical and if any experimental realisations already exist.

\begin{figure}
\caption{Magnetic structures giving rise to the magnetoelectric effect resulting from the action of the OP(a,b,c$\mid$d,e,f) transforming as mX$_{3}^{+}$ $\oplus$ mX$_{1}^{-}$, shown along the high symmetry directions OP(a,0,0$\mid$d,0,0), OP(a,a,0$\mid$d,d,0), OP(a,$\bar{a}$,a$\mid$d,$\bar{d}$,d).}
\includegraphics{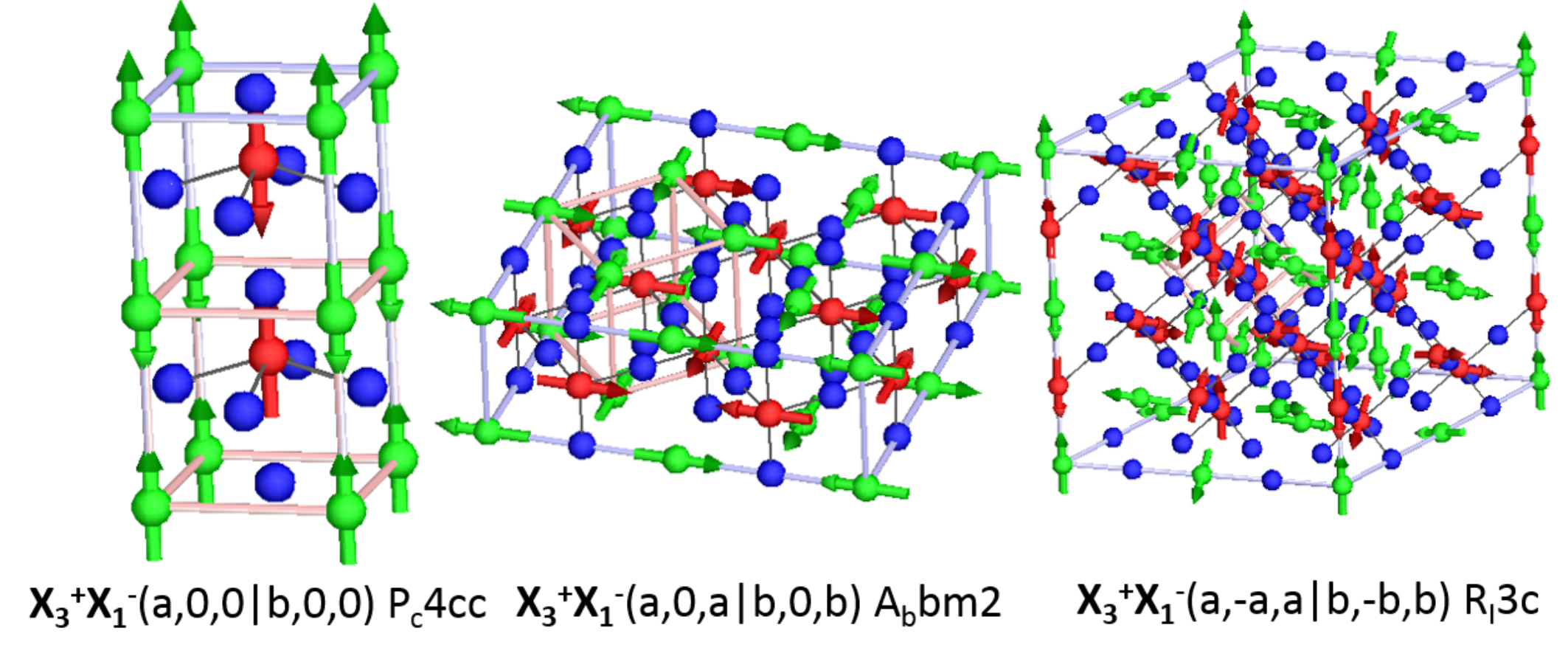}
\label{MagneticX}
\end{figure}

For mX$_{3}^{+}$ $\oplus$ mX$_{1}^{-}$, the order parameter is six dimensional OP(a;b;c$\mid$d;e;f), and the different choices of OPD result in a total of 22 possible isotropy subgroups. Only a subset of these in which the condition for conserving crystal momentum is satisfied at a linear term in polarisation, have broken inversion symmetry. While many of these lead to polar space groups, some only result in piezoelectric couplings. In these cases application of strain (either external or internal from ferroelastic distortions) will produce the desired polar ground state. These correspond to OPDs of OP(a;0;0$\mid$d;0;0), OP(a;0;a$\mid$d;0;d), OP(a;$\bar{a}$;a$\mid$d;$\bar{d}$;d), see Figure \ref{MagneticX}.  Of these only OP(a;0;0$\mid$d;0;0) represents a single k-active and collinear solution, and we shall focus on this for the rest of our discussion.  The isotropy group is $P_{c}4cc$ with basis=[(1,0,0),(0,0,1),(0,-2,0)] + (0,0,0) and SOPs $\Gamma_{4}^{-}$ (polar mode) and $\Gamma_{3}^{+}$ (tetragonal strain).  This OPD corresponds to the magnetic moments aligned parallel to the propagation vector on both A and B-sites.

\begin{figure}
\caption{Energy vs Polar mode ($\Gamma_{4}^{-}$ OP(0,h,0)) magnitude for AFM (mX$_3$$^+$ mX$_{1}^{-}$ OP(a,0,0$\mid$d,0,0)) and FM (m$\Gamma_{4}^{+}$) ordering. The inset illustrates the linear behaviour around the origin. The amplitude of the $\Gamma_{4}^{-}$ mode is determined by summing the displacements of all the atoms in the unit cell and presented as a percentage with respect to the ground state amplitude of the AFM phase. In both AFM and FM phases the energy shown is with respect to the structure with zero magnitude of $\Gamma_{4}^{-}$.}
\includegraphics[scale=0.5]{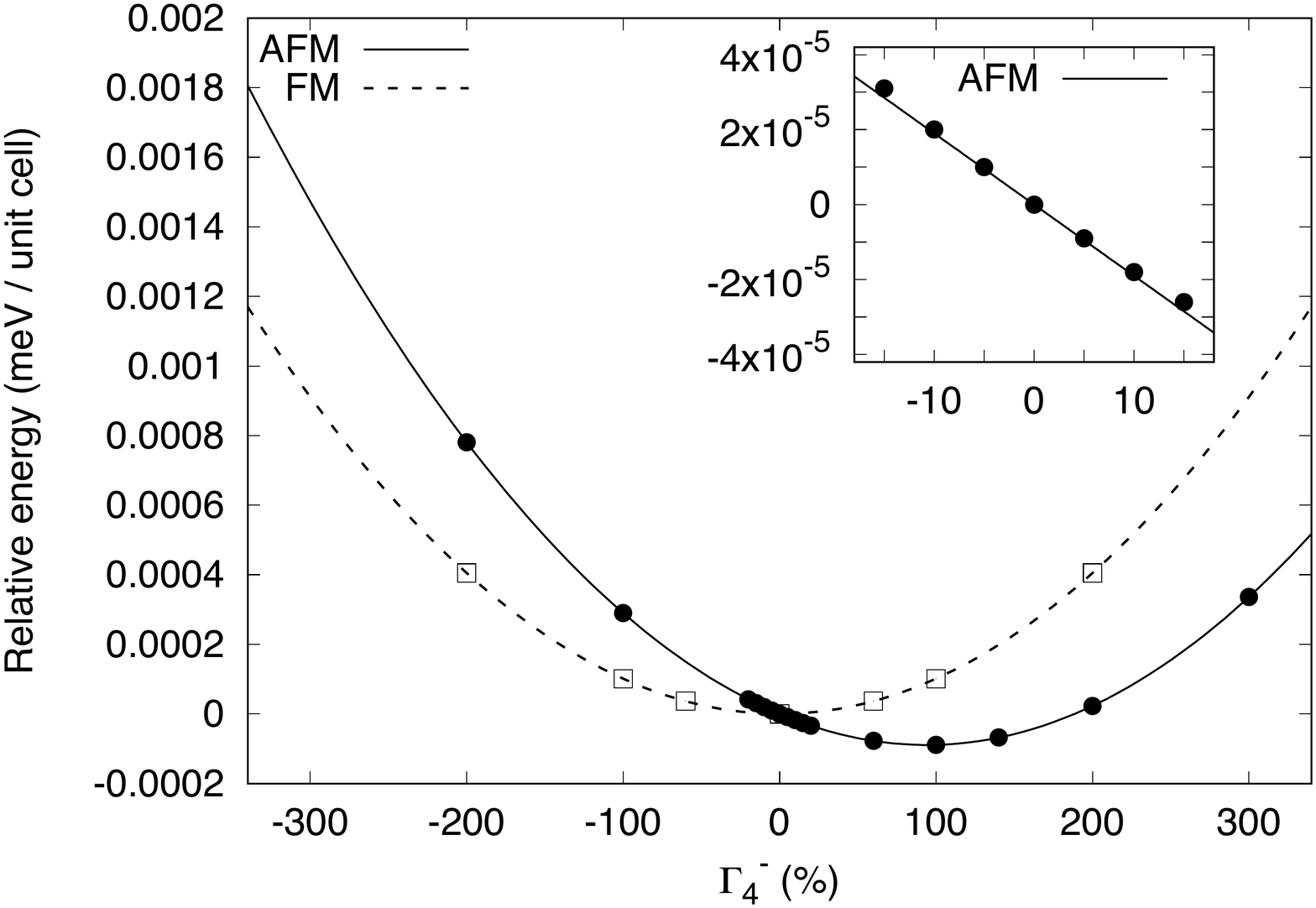}
\label{DFTX}
\end{figure}

To illustrate that our symmetry arguments can be used to identify improper ferroelectric couplings, we perform the following computational experiment.  
Density functional theory calculations using the $VASP$ code \cite{Kresse1993,Kresse1996} (version 5.4.1) were executed on a hypothetical cubic GdFeO$_3$ structure in which the unit cell parameter (the only degree of freedom), was fixed at $a$ =3.65 \AA ~.
This contracted unit cell was to insure that no polar instability existed in the phonon dispersion curve (in the ferromagnetic (FM) state, or with spin-orbit coupling turned off), such that any later appearance of $\Gamma_{4}^{-}$ (with spin-orbit coupling turned on) could be identified as arising through improper, rather than proper, ferroelectricity. 
This is illustrated in Figure \ref{DFTX} where the polar mode (GM4- OP (0,h,0)) is condensed with different amplitudes in the FM phase (m$\Gamma_{4}^{+}$) to give the expected single well potential centred at zero.  
%This indicates that our structure with fixed lattice parameters is at a minimum in total energy when the polar displacements are zero.   
We used the GGA PBEsol exchange correlation functional \cite{Perdew2008}, and PAW pseudopotentials (PBE functional, version 5.2) with the following valence electron configurations: $5s^25p^66s^24f^8$ (Gd), $3p^64s^23d^6$ (Fe) and $2s^22p^4$ (O).
An on-site Coulomb repulsion U \cite{Liechtenstein1995} was taken as 4 ev for the Gd f-electrons and 8ev for the Fe 3d-electrons, which further stiffened $\Gamma_{4}^{-}$, whilst keeping the system insulating. 
A plane wave cut-off of 900 eV and a 6 $\times$ 6 $\times$ 6 k-grid with respect to the cubic cell was employed.  

We then repeat these calculations with magnetic moments fixed on the A and B-sites that transform according to the irreps mX$_{3}^{+}$ and mX$_{1}^{-}$ (OP(a,0,0$\mid$d,0,0)).  
As evident in Figure \ref{DFTX}, the potential shifts away from having a minimum at zero (green line) prior to the magnetic interactions being switched on to a position where the minimum energy is at a finite value of the polar mode. This linear trend of the energy at the origin (inset Figure \ref{DFTX}) is indicative of an improper ferroelectric coupling term between mX$_{3}^{+}$, mX$_{1}^{-}$ and $\Gamma_{4}^{-}$.
We calculated the polarisation after full ionic relaxation to be 4.88 $\mu$C/cm$^2$, which we believe to be one of the largest reported amongst spin-induced ferroelectrics, suggesting a strong tri-linear coupling with this magnetic order.   
We compare this number to the purely electronic contribution to the polarisation calculated with the ions fixed in the high symmetry $Pm\bar{3}m$ positions, 0.07 $\mu$C/cm$^2$.
This suggests the total polarisation of 4.88 $\mu$C/cm$^2$ is predominantly of ionic origin, which is also suggested by the reasonably large cation-anion off-centering in the ground state structure (0.02 \AA).  Now that we have used these DFT calculations to illustrate our ideas, we will discuss the remaining magnetoelectric couplings based on symmetry arguments alone. 

\begin{figure}
\caption{Magnetic structures giving rise to the magnetoelectric effect resulting from the action of the OP(a,b,c$\mid$d,e,f) transforming as mX$_{5}^{+}$ $\oplus$ mX$_{5}^{-}$, shown along the high symmetry directions OP(a,a;0,0;0,0$\mid$d,$\bar{d}$;0,0;0,0), OP(a,a;0,0;0,0$\mid$$\bar{d}$,$\bar{d}$;0,0;0,0), OP(0,a;0,0;0,0$\mid$0,d,0,0;0,0).}
\includegraphics{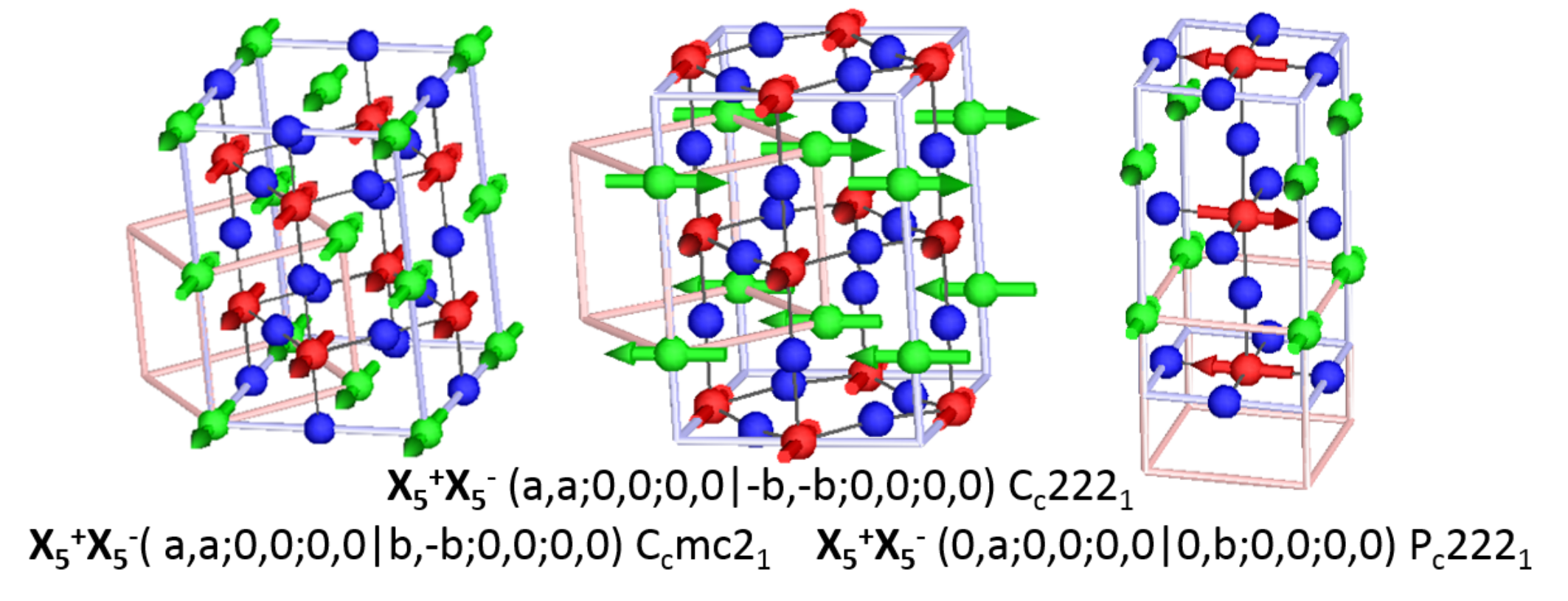}
\label{MagneticX2}
\end{figure}

For mX$_{5}^{+}$ $\oplus$ mX$_{5}^{-}$, the order parameter is now 12-dimensional OP(a,b;c,d;e,f$\mid$h,i;g,k;l,m). The representative (single k-active ) OPD which meet the criteria for zero crystal momentum transfer are however of the form OP(a,b;0,0;0,0$\mid$h,i;0,0;0,0).  We do not consider OP with multiple k-actives as in general this will always induce SOPs transforming as  M or R-point irreps, which are already covered in our previous analysis. We note here that we are not saying that these will correspond to physically equivalent examples, only that we can be sure that we have already consider the cases where linear terms in polarisation will also be present in the free energy expansions.  Hence the representative high symmetry examples given in Figure \ref{MagneticX2} are OP(a,a,0,0,0,0$\mid$h,$\bar{h}$,0,0,0,0) and OP(0,a,0,0,0,0$\mid \bar{h}$,0,0,0,0,0).  We do not explicitly consider mX$_{3}^{+}$ $\oplus$ mX$_{5}^{-}$ or mX$_{5}^{+}$ $\oplus$ mX$_{1}^{-}$ here as these represent magnetic structures in which the spins on the A-site and B-site are non-colinear with each other, which we believe to be less physically likely than the remaining examples that we have already discussed.

\begin{figure}
\caption{Collinear magnetic structures resulting from the action of the OP(a,b,c$\mid$d,e,f) transforming as mR$_{4}^{+}$ $\oplus$ mR$_{5}^{-}$, shown along the high symmetry directions OP(a,0,0$\mid$d,0,0), OP(a,a,0$\mid$d,d,0), OP(a,a,a$\mid$d,d,d).}
\includegraphics{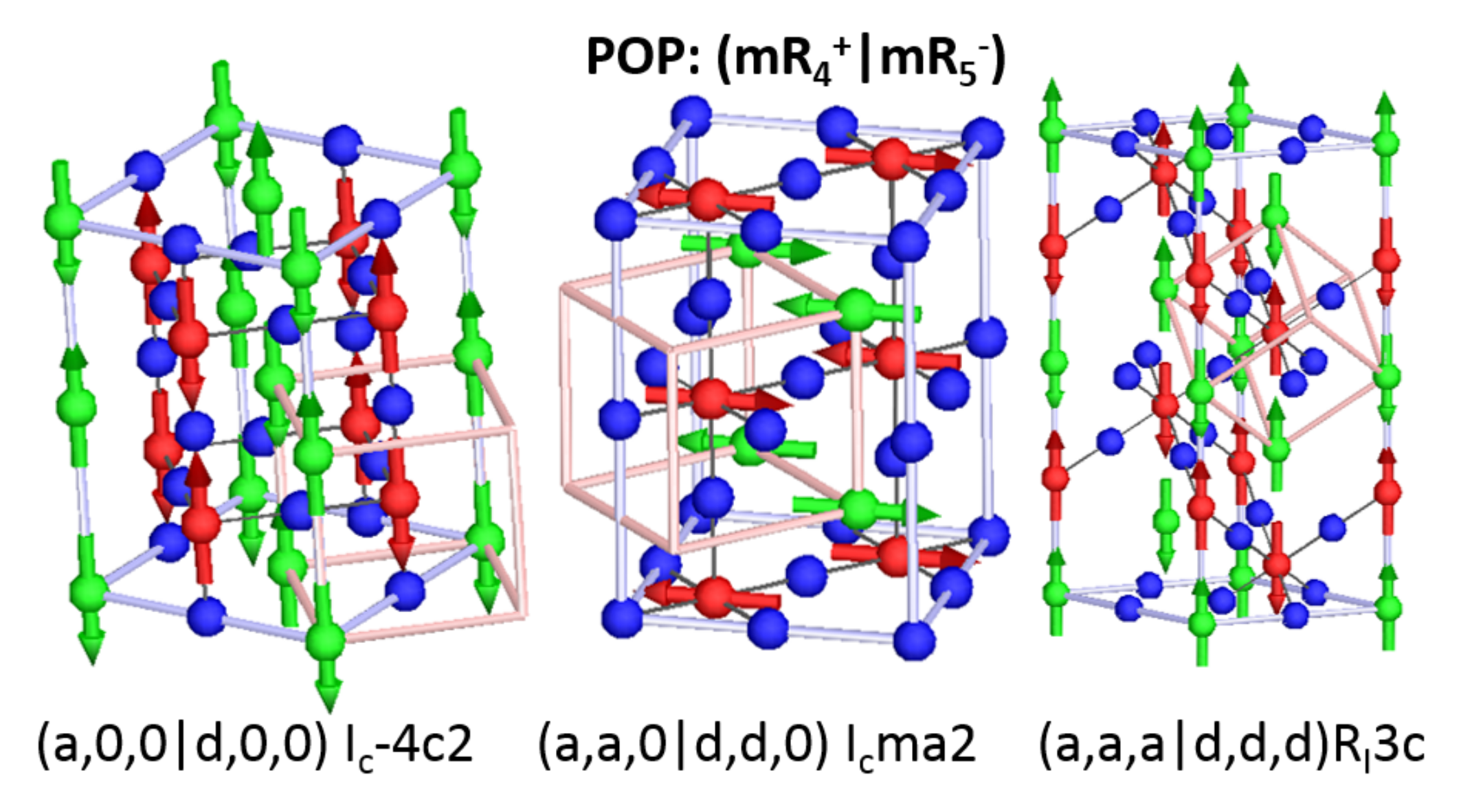}
\label{colinearR}
\end{figure}

For mR$_{4}^{+}$ $\oplus$ mR$_{5}^{-}$, the resulting order parameter is OP(a,b,c$\mid$d,e,f).  There are 14 possible order parameter directions that result in unique space group, basis and origin combinations with respect to the parent structure.  All other possible OPD correspond to twin domains of these 14 possibilities.  Of these 14 OPD, we consider here three: OP(a,0,0$\mid$d,0,0), $I_c\bar{4}c2$, basis=[(-1,-1,0),(1,-1,0),(0,0,2)] + (0,0,0); OP(a,a,0$\mid$d,d,0), $I_cma2$, basis=[(1,0,1),(1,0,-1),(0,2,0)] + (0,$\frac{1}{2}$0,-$\frac{1}{2}$0); OP(a,a,a$\mid$d,d,d) $R_I3c$, basis=[(1,0,-1),(0,-1,1),(-2,-2,-2)] + (0,0,0), which correspond to collinear magnetic structures shown in Figure~\ref{colinearR}.  Any of the other lower symmetry collinear magnetic structures may be constructed through linear combinations of these three OPD. For the polar space groups ($I_cma2$ and $R_I3c$) a SOP transforming as $\Gamma_{4}^{-}$ is always active. The only other SOPs are strain. A strategy for stabilising this ground state structure therefore, in addition to designing AFM nearest neighbour interaction in the system, is to epitaxially pre-strain the sample in a manner that stabilises terms in the free energy that will also occur at the even order.

$I_c\bar{4}c2$ (Figure~\ref{colinearR}, left) on the other hand, although it has no inversion symmetry, is only piezoelectric.  Indeed, it was very recently demonstrated~\cite{Zhao2017} from a combination of first principle calculations and group theoretical analysis that the rare earth Gadolinium chromates and ferrites with co-linear G-type order on A and B-site along the pseudo-cubic axes lead to a piezoelectric space group.  Sheer strain along the [110]-type lattice directions was found to be needed to create a polarisation through the piezoelectric effect, consistent with the piezoelectric point group. Our analysis shows an alternative route in which polarisation emerges directly, provided that the spins align along the orthorhombic or rhombohedral type axes as in the cases discussed above of OP(a,a,0$\mid$d,d,0) and OP(a,a,a$\mid$d,d,d).  We note also here that the possible observation of weak ferroelectric polarisation, which is reported in A and B-site lattices in which sub-lattice moments along 100-type directions are perpendicular to each other,~\cite{Zhao2017} may be understood in the frame work of the SOPs analysis that we have presented above.  We find that $\Gamma_{4}^{-}$ arises directly as a consequence of this kind of magnetic ordering (OP(a,0,0$\mid$,0,0,d)) with the magnetic space group being $F_Smm2$  (basis=[(0,2,0),(0,0,2),(2,0,0)], origin=(1/2,1/2,0)). 

An experimental example of where magnetoelectric properties arise from G-type ordering on the A and B-sties can be found in the literature for the 134-perovskite LaMn$_3$Cr$_4$O$_{12}$~\cite{Wang2015}.  This distorted perovskite structure has the additional structural orderings that can be described as M$_{1}^{+}$(a;a;a) (1:3 cation ordering) and M$_{2}^{+}$(a;a;a) (octahedral rotation). However, the observed magnetoelectric effect, that only occurs below both the B-site and the A-site ordering temperature, can be understood in terms of our present results by considering only OP mR$_{4}^{+}$ $\oplus$ mR$_{5}^{-}$ with OP(a,a,a$\mid$d,d,d)  (Figure~\ref{colinearR}, right), meaning that the magnetoelectric ground state structure has rhombohedral lattice symmetry and arises solely as a consequence of the magnetic ordering on both sites.

\section{Fourth order magnetoelectric couplings in AFM systems}

An exhaustive list of  fourth order couplings in polarisation and zone boundary irreps is given in Table \ref{fourthorder}.  There are naturally a large number of these, and we will restrict our more detailed discussion to those which are the most physically reasonable and likely to produce the strongest couplings at the highest ordering temperatures.  Because of this, we will no longer consider magnetic ordering on the A-site which in general only supports rare earth ions or non-magnetic cations.  Notable exceptions to this are the perovskite MnVO$_{3}$, but where the magnetic ordering temperatures remain low \cite{Markkula2011}, and some highly distorted AA'$_3$B$_4$O$_{12}$ quadruple perovskites that we will not discuss here.

Considering only B-site magnetism we are left with the following time-odd superposition of irreps to consider: mM$_{2,5}^{+}$ $\oplus$ mX$_{1,5}^{-}$; mR$_{5}^{-}$ $\oplus$ mX$_{1,5}^{-}$; mR$_{5}^{-}$ $\oplus$ mM$_{2,5}^{+}$.  In order to close the "momentum triangle" these will now be respectively superposed with the following time-even irreps:   R$_{1,5}^{+}$, M$_{3,5}^{-}$ and X$_{1}^{+}$, to produce an order parameter that transforms as time-even, inversion odd and has a crystal momentum transfer of zero (see Tables 3, 4 and 5).  The relevant structural degrees of freedom (Table \ref{ingredients}) to consider are hence, cation/anion order (X$_{1}^{+}$, R$_{1,5}^{+}$) and anti-polar displacements (M$_{3,5}^{-}$).  Notably, octahedral tilts or Jahn-Teller modes do not appear in this list and hence can not form part of such a design strategy.

\begin{figure}
\caption{Magnetic structures giving rise to the magnetoelectric effect resulting from the action of the OP(a;b;c$\mid$d;e;f$\mid$g,h,i) transforming as X$_{1}^{+}$ $\oplus$ mM$_{2}^{+}$ $\oplus$ mR$_{5}^{-}$, shown along the high symmetry directions indicated. A-site cation ordering is indicated by white and black spheres.}
\includegraphics{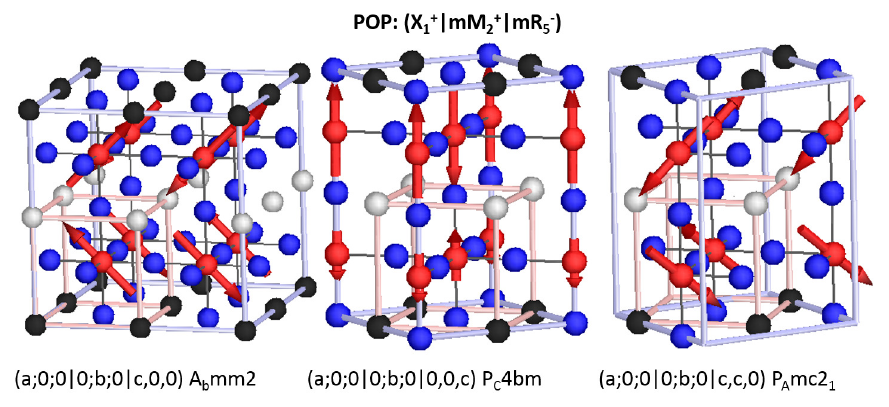}
\label{X1mM2mR5}
\end{figure}

\begin{figure}
\caption{Magnetic structures giving rise to the magnetoelectric effect resulting from the action of the OP(a;b;c$\mid$d,e;f,g;h,i$\mid$j,k,l) transforming as X$_{1}^{+}$ $\oplus$ mM$_{5}^{+}$ $\oplus$ mR$_{5}^{-}$, shown along the high symmetry directions indicated.  A-site cation ordering is indicated by white and black spheres.}
\includegraphics{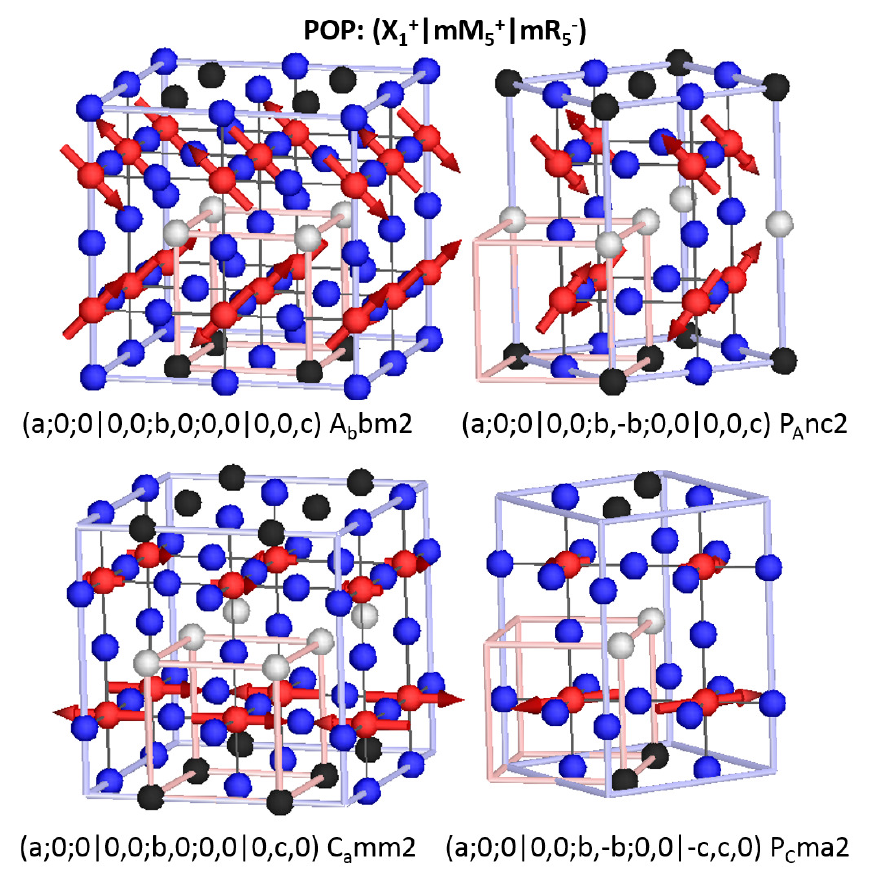}
\label{X1mM5mR5}
\end{figure}

For X$_{1}^{+}$ $\oplus$ mM$_{2,5}^{+}$ $\oplus$ mR$_{5}^{-}$ we give some possible examples of several magnetic structures in Figure \ref{X1mM2mR5} and \ref{X1mM5mR5} corresponding to A-site ordered double perovskites with striped type ( X$_{1}^{+}$) arrangements of cations, such as is commonly found experimentally for cations of substantially different sizes \cite{King2010}.  Some of these compounds are already known to be improper ferroelectric~\cite{Zuo2017} on account of couplings between the layering and octahedral tilt modes, as discussed in the previous section.   

The possible high symmetry OPD for superposed irreps X$_{1}^{+}$(0,$\frac{1}{2}$,0) $\oplus$ mM$_{2}^{+}$($\frac{1}{2}$,$\frac{1}{2}$,0) $\oplus$ mR$_{5}^{-}$($\frac{1}{2}$,$\frac{1}{2}$,$\frac{1}{2}$) are: \\
(a;0;0$\mid$0;b;0$\mid$c,0,0); (a;0;0$\mid$0;b;0$\mid$0,0,c); (a;0;0$\mid$0;b;0$\mid$c,c,0); 

Conservation of crystal momentum criteria that we have imposed here, dictates the relative OPD of the X and M components (k-actives).  The three structures listed above and shown in Figure \ref{X1mM2mR5} only differ in the OPD with respect to the mR$_{5}^{-}$ irrep, producing two non-collinear magnetic structures and one which has a spin-density-wave.  In the case of the non-collinear magnetic structures, the direction of P is parallel to both the cation order planes and the magnetic moment canting direction.  For the spin-density-wave structure the polarisation vector is perpendicular to the cation ordering planes.  Spin-density-wave magnetic structures are in general less common, but we note that X-point order of two magnetically active cations (at the B-site) with different magnetic moments could be a way to achieve this.

For X$_{1}^{+}$ $\oplus$ mM$_{5}^{+}$ $\oplus$ mR$_{5}^{-}$ as mM$_{5}^{+}$ is a higher dimensional irrep than M$_{2}^{+}$, there are now a larger number of OPD possibilities:\\
(a;0;0$\mid$0,0;b,0;0,0$\mid$c,0,0); (a;0;0$\mid$0,0;b,0;0,0$\mid$0,0,c); (a;0;0$\mid$0,0;b,0;0,0$\mid$0,c,0); \\(a;0;0$\mid$0,0;b,-b;0,0$\mid$0,0,c); (a;0;0$\mid$0,0;b,-b;0,0$\mid$c,c,0); (a;0;0$\mid$0,0;b,-b;0,0$\mid$-c,c,0)

However, this time several of these high symmetry order parameters give rise to piezoelectric but non-polar space groups ((a;0;0$\mid$0,0;b,0;0,0$\mid$c,0,0) $C222$ and (a;0;0$\mid$0,0;b,-b;0,0$\mid$c,c,0)  $P222_{1}$). Although not ferroelectric, the inclusion of any further POP either as an internal or external strain field will drive a ferroelectric ground state in these systems. Figure \ref{X1mM5mR5} shows the representative high symmetry OPD direction resulting in polar structure.  Similarly for the discussion above, P is parallel and perpendicular to cation ordering for constant moment and spin-density-wave magnetic structures respectively. 

\begin{figure}
\caption{Magnetic structures giving rise to the magnetoelectric effect resulting from the action of the OP as shown along the high symmetry directions indicated.  Anion ordering is indicated by blue spheres of differing sizes.}
\includegraphics{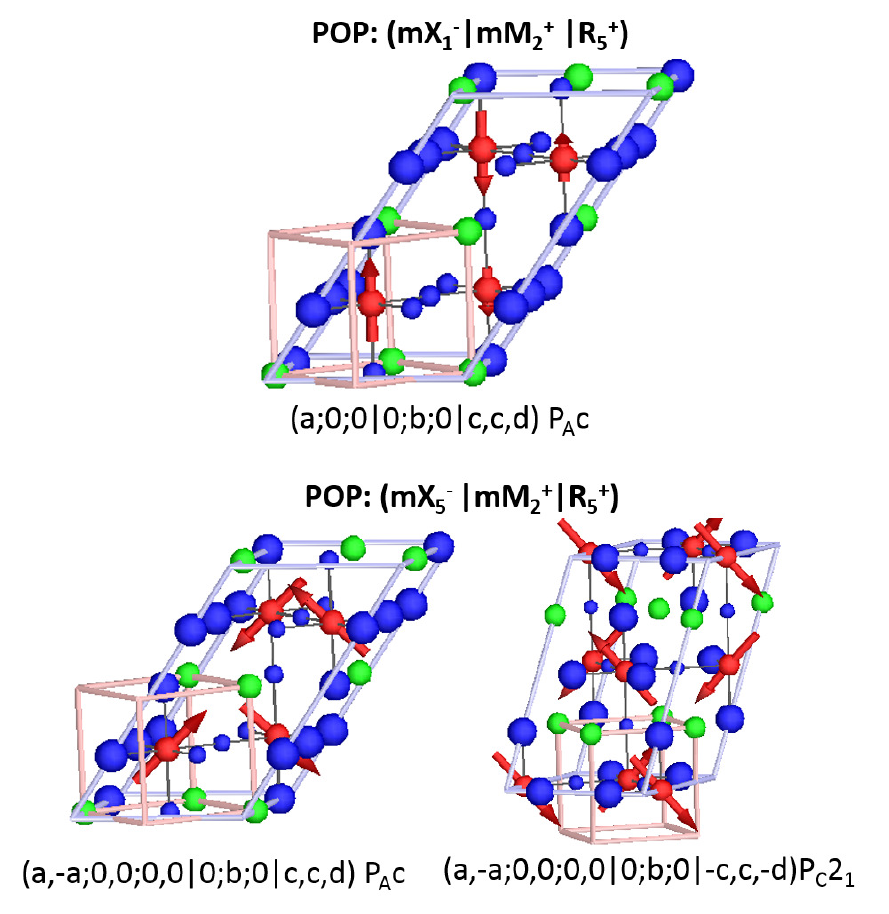}
\label{mXmM2R5}
\end{figure}

For, mX$_{1,5}^{-}$ $\oplus$ mM$_{2,5}^{+}$ $\oplus$ R$_{5}^{+}$, in which R$_{5}^{+}$ could correspond to anion-order, the cis-ordering of N for O substitution in oxynitride ABO$_{3-x}$N$_{x}$, perovskite \cite{Yang2011} represents an experimental realisation of this.  For $x$ = 1.5  , this would correspond to a checker-board anion order, and hence we consider R$_{5}^{+}$(a,a,a) (or the closest high symmetry equivalent OPD) in the following analysis.  As a POP transforming as R$_{5}^{+}$(a,a,a) always has a SOP transforming as R$_{1}^{+}$(a), this analysis also turns out to be equivalent to looking at rock-salt ordering on A-site cation, although we note that such ordering is not particularly common. mX$_{1}^{-}$(0,$\frac{1}{2}$,0) mM$_{2}^{+}$($\frac{1}{2}$,$\frac{1}{2}$,0) R$_{5}^{+}$($\frac{1}{2}$,$\frac{1}{2}$,$\frac{1}{2}$), with an OPD of (a;0;0$\mid$0;b;0$\mid$c,c,d), corresponds to a spin-density-wave collinear magnetic structure, where P is in the plane of the magnetic moment directions (Figure \ref{mXmM2R5}).  The constant magnitude spin canted magnetic structures (mX$_{5}^{-}$ $\oplus$ mM$_{2}^{+}$ $\oplus$ R$_{5}^{+}$ (a;0;0$\mid$0;b;0$\mid$c,c,d), Figure \ref{mXmM2R5}) on the other hand lead to polarisations that are found to be both perpendicular and parallel to the magnetic moment alignment.
 
\begin{figure}
\caption{Magnetic structures giving rise to the magnetoelectric effect resulting from the action of the OP as shown along the high symmetry directions indicated. Anion ordering is indicated by blue spheres of differing sizes.}
\includegraphics{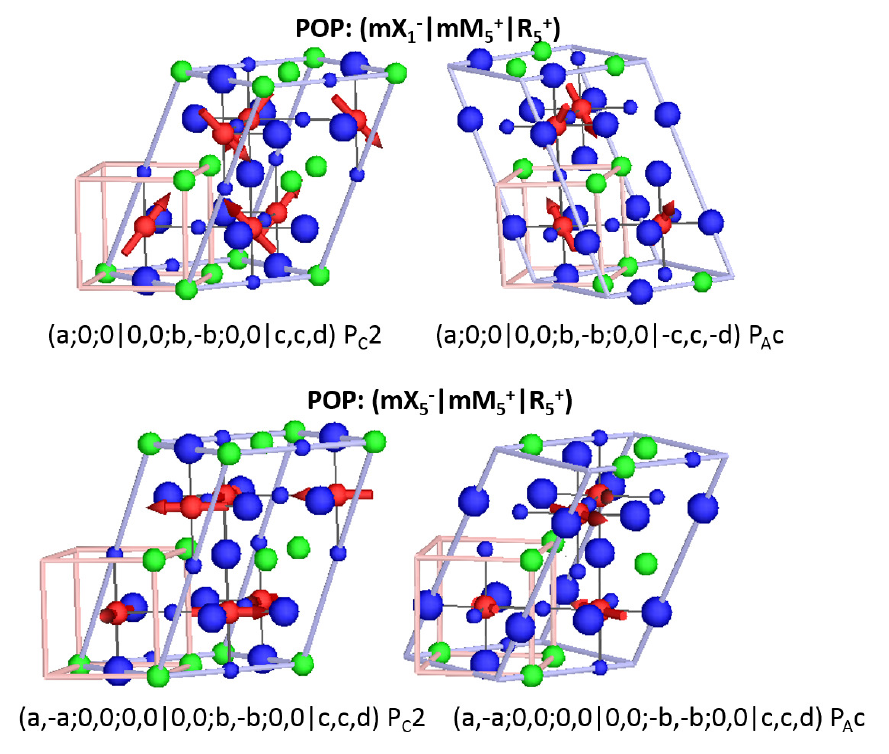}
\label{mXmM5R5}
\end{figure}

\begin{figure}
\caption{Magnetic structures giving rise to the magnetoelectric effect resulting from the action of the OP as shown along the high symmetry directions indicated.}
\includegraphics{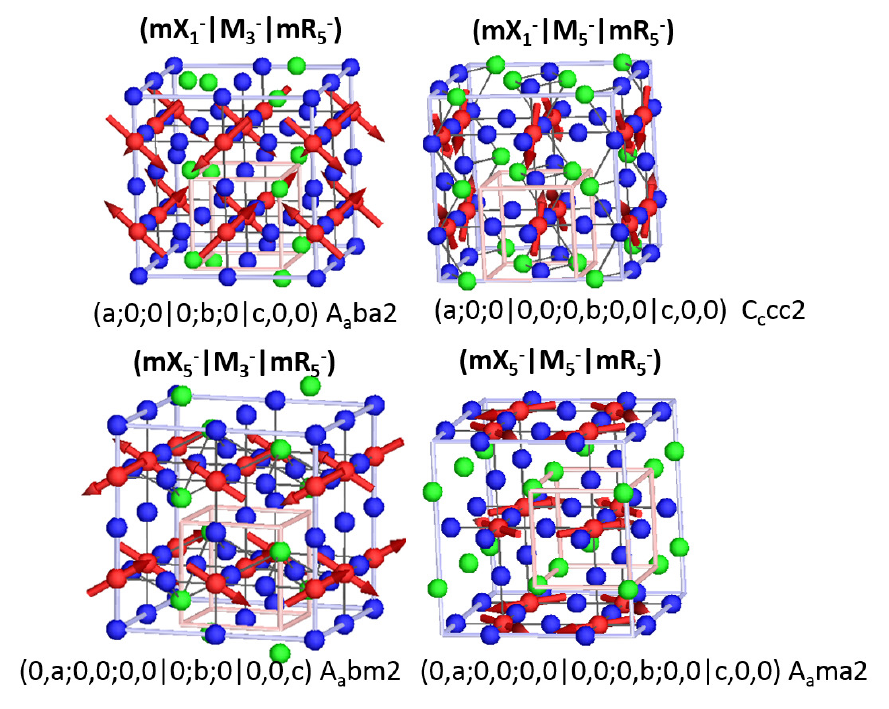}
\label{mXMmR}
\end{figure}

We will not consider the remaining possible couplings for mX$_{1}^{-}$ $\oplus$ mM$_{2}^{+}$ $\oplus$ R$_{5}^{+}$, mX$^{-}$ $\oplus$ mM$_{5}^{+}$ $\oplus$ R$_{5}^{+}$ and mX$_{1}^{-}$ $\oplus$ M$^{-}$ $\oplus$ mR$_{5}^{-}$ explicitly here but they are tabulated in Table \ref{SOPsR} and representative figures are given in Figure \ref{mXmM2R5}, \ref{mXmM5R5} and \ref{mXMmR}.

\section{Stabilising wFM and Magnetoelectric effects in non-ferroelectrics}

For multiferroics to be useful in data storage applications, it is likely that they will need to have a ferromagnetically ordered and switchable component.  Following the design strategy above we wish to engineer a trilinear term in the free energy expansion of the form A B wFM.  Since wFM transforms as m$\Gamma_{4}^{+}$ which is time odd, parity even, and has crystal momentum of zero, the constraints on A and B are as follows:

[A] = [B] \\
A$\cdot$B is parity even \\
Hence possible trilinear terms with wFM need involve OPs that transform as:\\
mR$^{-}$ $\oplus$  R$^{-}$; mR$^{+}$ $\oplus$  R$^{+}$; mM$^{+}$ $\oplus$  M$^{+}$; mX$^{-}$ $\oplus$  X$^{-}$; mX$^{+}$ $\oplus$  X$^{+}$;.

Taking mR$_{5}^{-}$ (B-site magnetic order) with R$_{5}^{-}$ (octahedral tilting) as an example, G-type magnetic ordering on the B-sites with moments along the c-axis with out of phase octahedral rotations leads to the magnetic space group Im'm'a, (basis=[(1,1,0),(0,0,2),(1,-1,0)]+ (0,0,$\frac{1}{2}$)) which has m$\Gamma_{4}^{+}$ (wFM) as a SOP.  Indeed, we believe this is the frame work under which the theoretically predicted wFM  in Gd Cr/Fe perovskites \cite{Zhao2017,Tokunaga2009} can be easily understood, and is an example in which the B-O-B exchange angle is allowed to deviate from 180 degrees by a symmetry breaking event allowing spin canting to occur via the Dzyaloshinsky-Moriya interaction.

At the M-point the above analysis can also be applied.  C-type B-site magnetic ordering along the [001]-axis (mM$_{2}^{+}$(a;0;0)), with in phase octahedral tilts perpendicular to this (M$_{2}^{+}$(a;0;0), $a^0a^0c^+$), actually leads to a piezomagnetic space group $P4/mbm$ (basis=[(1,-1,0),(1,1,0),(0,0,1)] +($\frac{1}{2}$,-$\frac{1}{2}$,0)). Application of a orthorhombic type strain ($\Gamma_{5}^{+}$) for example leads to the occurrence of wFM (m$\Gamma_{4}^{+}$).  Similarly distortions transforming as X$_{5}^{-}$ and magnetic moments as mX$_{1}^{-}$ will produce wFM (e.g. $Cm'cm'$, basis=[(1,0,1),(1,0,-1),(0,2,0)] + (0,$\frac{1}{2}$,0)).

Another magnetoelectric effect worth considering is where P is induced by the application of external magnetic field which may be described as transforming as m$\Gamma_{4}^{+}$ or conversely wFM is induced by the application of an external electric field ($\Gamma_{4}^{-}$).  For this we must look at terms involving two zone boundary irreps like [M] + [S] = [0 0 0], where M is time odd (magnetic) and S is time even (structural), and M$\cdot$S is inversion odd.  Application of an electric field ($\Gamma_{4}^{-}$) should then give a fourth order term in the free energy expansion of the form M S P wFM.  A realisations of this is mX$_{1,5}^{-}$ $\oplus$ X$_{1}^{+}$ $\oplus$ P, to give wFM.  Finally it is worth pointing out that all systems that are both piezoelectric and piezomagnetic will be magnetoelectric, as application of either an external magnetic or electric field will generate a strain field that mediates a coupling between the two phenomena. 
%CeBaMn2O6?

\section{Putting it all Together}

\begin{figure}
\caption{A scheme for including third order couplings term in to the free energy expansion involving the order parameter related to weak ferromagnetic spin canting and ferroelectric polarisation.}
\includegraphics{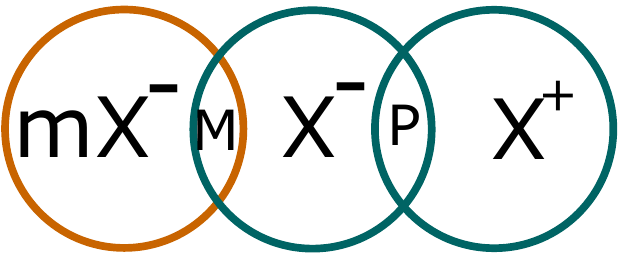}
\label{wFmschemethird}
\end{figure}

The ultimate goal of course is to have a magnetoelectric in which ferromagnetism is coupled to ferroelectricity. To achieve the strongest such coupling, we envisage first a scenario with two trilinear terms in P and wFM, with one codependent order parameter (see Figure \ref{wFmschemethird}).  For example: (1) X$_{1}^{+}$ $\oplus$ X$_{5}^{-}$ $\oplus$ P with mX$_{1}^{-}$ $\oplus$ X$_{5}^{-}$ $\oplus$ wFM. Assuming X$_{1}^{+}$ represents cation order and may not be reversed, then the reversal of the sign of P would necessitate a reversal of X$_{5}^{-}$. This, in turn, would necessitate a switching of the magnetic structure which most likely would proceed via a reversal of the direction of the wFM. (2) M$_{5}^{-}$ $\oplus$ M$_{2}^{+}$ $\oplus$ P and mM$_{5}^{-}$ $\oplus$ M$_{2}^{+}$ $\oplus$ wFM, Taking M$_{5}^{-}$ as anion ordering, then a reversal of P would proceed via reversal of the octahedral rotations (M$_{2}^{+}$) necessitating a reversal of either mM$_{5}^{-}$ or wFM, the later being more likely. (3) R$_{1}^{+}$ $\oplus$ R$_{2}^{-}$ $\oplus$ P and mR$_{5}^{-}$ $\oplus$ R$_{2}^{-}$ $\oplus$ wFM, taking R$_{1}^{+}$ as A-site rock salt cation ordering, a reversal of P would imply a switching of R$_{2}^{-}$ which could represent B-site charge ordering

In fact, if one has an AA' layered double Perovskite (X$_{1}^{+}$) with the common M$_{2}^{+}$ and R$_{5}^{-}$  tilt pattern (Pnma like) no matter if you have  C(mM$_{5}^{+}$),G(mR$_{5}^{-}$) or A(mX$_{1}^{-}$) magnetic ordering (provided the spins are along certain directions), the ground state is ferroelectric and ferromagnetic with an indirect coupling between them.  Efforts should hence be focused on preparing A$^{+}$ A$^{'3+}$ and A$^{2+}$ A$^{'4+}$   layered double perovskites with Mn$^{4+}$ and Fe$^{3+}$ on the B-site, respectively, to achieve the strongest wFM moments and highest ordering temperatures.  

\begin{figure}
\caption{A scheme for including fourth order couplings term in to the free energy expansion involving the order parameter related to weak ferromagnetic spin canting and ferroelectric polarisation.}
\includegraphics{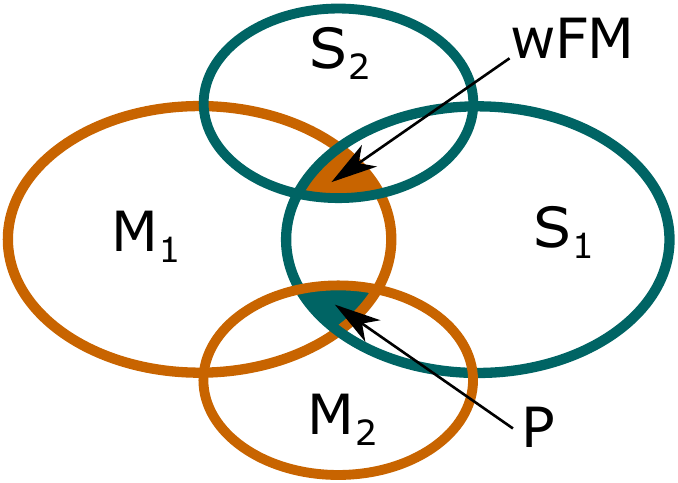}
\label{wFmscheme}
\end{figure}

Another scheme involving fourth order couplings gives a greater degree of flexibility.  Similar to the above, the idea here is to construct fourth order terms with wFM (m$\Gamma_{4}^{+}$) in.  As many as the OPs featuring in the wFM term at the fourth order should also feature in the fourth order term in P.  Figure \ref{wFmscheme} envisages one such possible coupling scheme by which an extra degree of freedom related to breaking structural symmetry (S$_2$) is introduced to the magnetoelectric couplings discussed above, and is equivalent to using antisymmetric (DM) arguments to design wFM.   The figure shows that it is possible to construct fourth order terms with at least two OP in common in both  P and wFM terms,  i.e. M$_1$ M$_2$ S$_1$ P and M$_1$ S$_1$ S$_2$ wFM.  Each fourth order term must individually conserve crystal momentum, time reversal and inversion symmetry. Hence the polar part, M$_1$ M$_2$ S$_1$, can be selected according to the analysis in the previous section, leaving the wFM part, M$_1$ S$_1$ S$_2$, to be decided on.  Since M$_1$ S$_1$ are fixed by the polar part, the only decision to be made is the nature of S$_2$.  We require that the crystal momentum of [S$_2$] equals the sum of the crystal momentum [M$_1$] + [M$_2$], and that parity with respect to inversion equal to the product of the parity of M$_1$$\cdot$S$_1$ (i.e. opposite to that of M$_2$).  
For example, with mM$_{2,5}^{+}$ (M$_1$), mR$_{5}^{-}$ (M$_2$) and X$_{1}^{+}$ (S$_1$), S$_2$ must be either R$_{1,5}^{+}$ or M$_{2,3,5}^{-}$.

Finally we note that, if one predisposes the system to certain distortions which are implied as SOPs in the above analysis, certain phases may be thermodynamically favoured over others.  This is an important part of controlling the relative order parameter directions which ultimately effect the higher order couplings that drive magnetoelectric properties.   We discuss now a few of the most promising candidates and propose some design strategies based on SOP analysis.  SOPs are listed in Table \ref{SOPsX}, \ref{SOPsM}~and \ref{SOPsR} for some fourth order couplings in ferroelectric polarisation. Any further distortion to the $Pm\bar{3}m$ aristotype that the system is predisposed to, which transform as irreps in this list, will act to stabilise one particular order parameter direction over another. Or put another way, at the harmonic order (quadratic phonon modes) all possible OPD are degenerate in energy.

The most obvious strategy is to pre-strain (transforming as $\Gamma_{3,5}^{+}$) the system through epitaxial growth.  Another strategy is to search through tables such as Table \ref{SOPsX}, \ref{SOPsM} and \ref{SOPsR}, to find irreps that correspond to the most commonly observed distortions in the perovskite phase, such as the octahedral rotations ( M$_{2}^{+}$ and R$_{5}^{-}$), Jahn-Teller distortions in systems with a degeneracy in their d-orbitals or indeed polar distortions themselves in $d^0$-systems.  In the ``undistorted'' perovskite structure these will corresponded to the lowest lying phonon modes (rigid unit modes in the cases of the octahedral rotation). Any energy penalty paid at the quadratic order will be kept low with respect to the trilinear terms that always act to lower the free energy, and therefore will drive a phase transition.   For example, for X$_{1}^{+}$ $\oplus$ mM$_{2}^{+}$ $\oplus$ mR$_{5}^{-}$, SOPs are strain, $\Gamma_{5}^{-}$, and X-point distortions. The OP(a;0;0$\mid$0;b;0$\mid$c,c,0) $P_{A}mc2_{1}$ is the most promising candidate, as the only SOP is anti-polar X5- distortions.  Therefore, in addition to striped cation ordering, cations which are susceptible to off-centre distortions should be chosen.  For mX$_{1}^{-}$ $\oplus$ M$_{5}^{-}$ $\oplus$ mR$_{5}^{-}$ or mX$_{1}^{-}$ $\oplus$ mM$_{2}^{+}$ $\oplus$ R$_{5}^{+}$, B-sites with a propensity to undergo charge (M4+ or R2-) and orbital order (M3+ or R3-) should be chosen.  A similar design strategy of selecting a system which is predisposed to certain SOPs may be adopted for stabilising wFM.

\section{Conclusion}

Using group theoretical means, we have enumerated the possible magnetoelectric couplings in the perovskite structure with respect to its aristotypical symmetry $Pm\bar{3}m$.  Our enumeration is complete up to the third order terms for zone-boundary magnetic structures, and for fourth order terms for B-site magnetism only.  Our results show that, for zone-boundary magnetic ordering, only magnetism on both A and B sites transforming either both as X-point or R-point irreps can produce a magnetoelectric coupling at the third order, which is illustrated with first principles calculations.  For magnetism on the B-site alone, then only fourth order terms can produce the desired effect.  We propose a design strategy based on POPs consisting of a superposition of three irreps one each from the X, M and R-point, chosen in such a way that crystal momentum is conserved, that two are time-odd and either one or all are inversion-odd.  These ideas are extended to a design strategy for weak ferromagnetism, which may then be coupled to the ferroelectric polarisation in a similar manner of the recently much discussed hybrid improper ferroelectric Ca$_3$Mn$_2$O$_7$.  Without a doubt, predicting and controlling physical properties arising from magnetic order will remain a challenging area for many years to come.  However, our systematic enumeration of coupling mechanisms along with secondary order parameters at least provides some direction for how this might ultimately be systematically achieved.

\appendix
\section{Notation used in order parameter directions}
We will be forming order parameters from up to three different zone boundary irreps, which we will denote by using the symbol for direct sum '$\oplus$', for example M$_{2}^{+}$ $\oplus$ R$_{5}^{-}$ represents an order parameter that has atomic displacements that transform as both irreps. By forming this OP we will effectively be conducting a thought experiment as to what would happen if the parent $Pm\bar{3}m$ became spontaneously unstable with respect to atomic displacements (in this case octahedral rotations) transforming as these irreps.  However, specifying these irreps alone does not capture how the displacements (or magnetic orderings) combine with respect to each other, and hence the associated isotropy subgroup.  To do this we need to describe the full OPD of the OP transforming as the specified irreps, and we follow the notation used in $ISODISRORT$\cite{Campbell2006}.  For M$_{2}^{+}$ $\oplus$ R$_{5}^{-}$, a OPD is OP(a;0;0$\mid$b,0,0), where "$\mid$" denotes a division between the OPD parts belonging to M$_{2}^{+}$ and R$_{5}^{-}$ respectively.  Semi-colons ";" denote divisions between different OPD resulting from the degeneracy of the propagation vector in $Pm\bar{3}m$.  At the M-point the possible k-actives are ($\frac{1}{2}$,$\frac{1}{2}$,0);($\frac{1}{2}$,0,$\frac{1}{2}$);(0,$\frac{1}{2}$,$\frac{1}{2}$), where the order respects the order of appearance in the OPD. Similarly for OP transforming as X-point irreps we will need to specify which k-actives of (0,$\frac{1}{2}$,0);($\frac{1}{2}$,0,0);(0,0,$\frac{1}{2}$) are in use. In general we will only form order parameters from one k-active per irrep (equivalent to using the small irrep only). However, the notation we give will always reflect the total number of possible k-actives.  At the R-point there is only one possible k-active, but in this case the irrep is multi-dimensional and, as in the case of $\Gamma_{4}^{-}$, this is specified through use of commas between the letters. The total dimension of the OP is hence a function of the number of superposed irreps, the degeneracy of the propagation vectors associated with any of the irreps, and the dimensionality of the small irreps themselves. All need to be fully specified along with the setting and space group of the parent to uniquely identify the isotropy subgroup.

\ack{Acknowledgements}

M.S.S acknowledges the Royal Commission for the Exhibition of 1851 and the Royal Society for fellowships. N.C.B. also acknowledges the Royal Commission for the Exhibition of 1851 for a fellowship and is also supported by an Imperial College Research Fellowship and a Royal Society Research grant. M.S.S. would like to acknowledge Hanna Bostr\"{o}m for useful discussions.
Calculations were performed on the Imperial College London high-performance computing facility. We are also grateful to the UK Materials and Molecular Modelling Hub for computational resources, which is partially funded by EPSRC (EP/ P020194/1).

% References are at the end of the document, between \begin{references}
% and \end{references} tags. Each reference is in a \reference entry.

%\begin{references}
\bibliographystyle{iucr}
\bibliography{bib5} 
%\end{references}

     %-------------------------------------------------------------------------
     % TABLES AND FIGURES SHOULD BE INSERTED AFTER THE MAIN BODY OF THE TEXT
     %-------------------------------------------------------------------------

     % Simple tables should use the tabular environment according to this
     % model

\begin{table}
\label{ingredients}
\caption{Ingredients for symmetry breaking in the Perovskite structure, classified in terms of transforming as irreps of the parent perovskite structure, with the A-site at the origin. (The corresponding irrep labels for the setting with the B-site at the origin are given in brackets)}
\begin{tabular}{lcccc}      % Alignment for each cell: l=left, c=center, r=right
Ingredient            & $\Gamma$               & X          & M          & R          \\
\hline
Strain                & $\Gamma_{3}^{+}$; $\Gamma_{5}^{+}$ & -        & -                  & -          \\
Cation order (A)      & -                     & X$_{1}^{+}$(X$_{3}^{-}$)              & M$_{1}^{+}$(M$_{4}^{+}$)           & R$_{1}^{+}$(R$_{2}^{-}$)        \\
Cation order (B)      & -                     & X$_{3}^{-}$(X$_{1}^{+}$)              & M$_{4}^{+}$(M$_{1}^{+}$)           & R$_{2}^{-}$(R$_{1}^{+}$)        \\
Anion order (O)       & -                      & X$_{1}^{+}$(X$_{3}^{-}$)             & M$_{4}^{+}$(M$_{1}^{+}$); M$_{5}^{-}$(M$_{5}^{-}$) & R$_{5}^{+}$(R$_{4}^{-}$)        \\
(Anti-)Polar (A)      & $\Gamma_{4}^{-}$      & X$_{3}^{-}$(X$_{1}^{+}$); X$_{5}^{-}$(X$_{5}^{+}$)    & M$_{3}^{-}$(M$_{2}^{-}$); M$_{5}^{-}$(M$_{5}^{-}$) & R$_{4}^{-}$(R$_{5}^{+}$)        \\
(Anti-)Polar (B)      & $\Gamma_{4}^{-}$      & X$_{1}^{+}$(X$_{3}^{-}$); X$_{5}^{+}$(X$_{5}^{+}$)    & M$_{2}^{-}$(M$_{3}^{-}$); M$_{5}^{-}$(M$_{5}^{-}$) & R$_{5}^{+}$(R$_{4}^{-}$)        \\
Jahn-Teller modes     & $\Gamma_{3}^{+}$      & X$_{3}^{-}$(X$_{1}^{+}$)              & M$_{3}^{+}$(M$_{2}^{+}$)           & R$_{3}^{-}$(R$_{3}^{+}$)        \\
Octahedral tilt modes & -                     & -                     & M$_{2}^{+}$(M$_{3}^{+}$)           & R$_{5}^{-}$(R$_{4}^{+}$)        \\
Magnetic order (A)    & m$\Gamma_{4}^{+}$     & mX$_{3}^{+}$(mX$_{1}^{-}$);mX$_{5}^{+}$(mX$_{5}^{-}$)             & mM$_{3}^{+}$(mM$_{2}^{+}$); mM$_{5}^{+}$(mM$_{5}^{+}$)         & mR$_{4}^{+}$(mR$_{5}^{-}$) \\
Magnetic order (B)    & m$\Gamma_{4}^{+}$     & mX$_{1}^{-}$(mX$_{3}^{+}$);mX$_{5}^{-}$(mX$_{5}^{+}$)             & mM$_{2}^{+}$(mM$_{3}^{+}$); mM$_{5}^{+}$(mM$_{5}^{+}$)         & mR$_{5}^{-}$(mR$_{4}^{+}$)

\end{tabular}
\end{table}

\begin{table}
\label{fourthorder}
\caption{Closing the "momentum triangle" - the possible fourth order magnetoelectric coupling terms.  Zeroth row and column correspond to two of the 4 coupling terms which are always time odd.  At the intersection of the rows and columns, a third-time even irrep is given with the fourth term always being P ($\Gamma_{4}^{-}$).}
\begin{tabular}{lcc}      % Alignment for each cell: l=left, c=center, r=right
-- & mX$_{3}^{+}$ \& mX$_{5}^{+}$ & mX$_{1}^{-}$ \& mX$_{5}^{-}$  \\
\hline
mM$_{2}^{+}$  & R$_{2}^{-}$; R$_{3}^{-}$; R$_{4}^{-}$; R$_{5}^{-}$  &  R$_{1}^{+}$; R$_{5}^{+}$   \\  
mM$_{3}^{+}$  & R$_{2}^{-}$; R$_{3}^{-}$; R$_{4}^{-}$; R$_{5}^{-}$  &  R$_{1}^{+}$; R$_{5}^{+}$   \\  
mM$_{5}^{+}$  & R$_{2}^{-}$; R$_{3}^{-}$; R$_{4}^{-}$; R$_{5}^{-}$  &  R$_{1}^{+}$; R$_{5}^{+}$   \\  
\end{tabular}

\begin{tabular}{lcc}      % Alignment for each cell: l=left, c=center, r=right
-- & mX$_{3}^{+}$ \& mX$_{5}^{+}$ & mX$_{1}^{-}$ \& mX$_{5}^{-}$  \\
\hline
mR$_{4}^{+}$  &   M$_{3}^{-}$; M$_{5}^{-}$     &     M$_{1}^{+}$; M$_{2}^{+}$; M$_{3}^{+}$; M$_{4}^{+}$      \\ 
mR$_{5}^{-}$  &   M$_{1}^{+}$; M$_{2}^{+}$; M$_{3}^{+}$; M$_{4}^{+}$     &    M$_{3}^{-}$; M$_{5}^{-}$      \\  
\end{tabular}

\begin{tabular}{lc}      % Alignment for each cell: l=left, c=center, r=right
-- &  mM$_{2}^{+}$ \& mM$_{3}^{+}$ \& mM$_{5}^{+}$   \\
\hline
mR$_{4}^{+}$  &    X$_{3}^{-}$; X$_{5}^{-}$     \\ 
mR$_{5}^{-}$  &    X$_{1}^{+}$                  \\  
\end{tabular}
\end{table}

\begin{table}
\label{SOPsX}
\caption{Structural SOPs of POPs indicated in the table. Polarisation $\Gamma_{4}^{-}$ is always a SOP.}
\begin{tabular}{lc} % Alignment for each cell: l=left, c=center, r=right
POP & SOPs \\
\hline\hline
(X$_1^+$$\mid$mM$_2^+$$\mid$mR$_5^-$) & - \\
(a;0;0$\mid$0;b;0$\mid$c,0,0) & $\Gamma_3^+$(a,b); $\Gamma_4^-$(0,0,a); $\Gamma_5^-$(a,0,0); X$_2^+$(a;0;0); X$_5^-$(a,0;0,0;0,0) \\
(a;0;0$\mid$0;b;0$\mid$0,0,c) & $\Gamma_3^+$(a,$\sqrt{3}\bar{a}$); $\Gamma_4^-$(0,a,0); X$_3^-$(a;0;0) \\
(a;0;0$\mid$0;b;0$\mid$c,c,0) & $\Gamma_3^+$(a,$\sqrt{3}\bar{a}$); $\Gamma_5^+$(0,0,a); $\Gamma_4^-$(a,0,a); $\Gamma_5^-$(a,$\bar{a}$,0); X$_5^-$(a,a;0,0;0,0)\\
\hline
(X$_1^+$$\mid$mM$_5^+$$\mid$mR$_5^-$) & -\\ 
(a;0;0$\mid$0,0;b,0;0,0$\mid$0,0,c) & $\Gamma_3^+$(a,b); $\Gamma_4^-$(a,0,0); $\Gamma_5^-$(0,a,0); X$_2^+$(a;0;0); X$_5^-$(0,a;0,0;0,0)\\
(a;0;0$\mid$0,0;b,0;0,0$\mid$0,c,0) & $\Gamma_3^+$(a,b); $\Gamma_4^-$(0,a,0); $\Gamma_5^-$(0,0,a); X$_2^+$(a;0;0); X$_3^-$(a;0;0) \\
(a;0;0$\mid$0,0;b,$\bar{b}$;0,0$\mid$0,0,c) & $\Gamma_3^+$(a,$\sqrt{3}\bar{a}$); $\Gamma_5^+$(0,0,a); $\Gamma_4^-$(a,0,$\bar{a}$); $\Gamma_5^-$(a,a,0); X$_5^-$(a,$\bar{a}$;0,0;0,0) \\
(a;0;0$\mid$0,0;b,$\bar{b}$;0,0$\mid$$\bar{c}$,c,0) & $\Gamma_3^+$(a,$\sqrt{3}\bar{a}$); $\Gamma_5^+$(0,0,a); $\Gamma_4^-$(0,a,0); X$_3^-$(a;0;0)\\
\hline\hline
\end{tabular}
\end{table}

\begin{table}
\label{SOPsM}
\caption{Structural SOPs of POPs indicated in the table. Polarisation, $\Gamma_{4}^{-}$ is always a SOP}
\begin{tabular}{lc} % Alignment for each cell: l=left, c=center, r=right
POP & SOPs \\
\hline\hline
(mX$_1^-$$\mid$M$_3^-$$\mid$mR$_5^-$) & - \\
(a;0;0$\mid$0;b;0$\mid$c,0,0) & $\Gamma_3^+$(a,b); $\Gamma_4^-$(a,0,0); $\Gamma_5^-$(0,a,0); M$_5^+$(0,0;a,0;0,0) \\
(a;0;0$\mid$0;b;0$\mid$c,c,0) & $\Gamma_3^+$(a,$\sqrt{3}\bar{a}$); $\Gamma_5^+$(0,0,a); $\Gamma_4^-$(a,0,a); $\Gamma_5^-$(a,$\bar{a}$,0); M$_5^+$(0,0;a,$\bar{a}$;0,0); M$_2^-$(0;a;0)\\
\hline 
(mX$_5^-$$\mid$M$_3^-$$\mid$mR$_5^-$) & - \\
(a,$\bar{a}$;0,0;0,0$\mid$0;b;0$\mid$0,0,c) & $\Gamma_3^+$(a,$\sqrt{3}\bar{a}$); $\Gamma_5^+$(0,0,a); $\Gamma_4^-$(a,0,a); $\Gamma_5^-$(a,$\bar{a}$,0); M$_5^+$(0,0;a,$\bar{a}$;0,0); M$_2^-$(0;a;0) \\
(0,a;0,0;0,0$\mid$0;b;0$\mid$0,0,c) & $\Gamma_3^+$(a,b); $\Gamma_4^-$(a,0,0); $\Gamma_5^-$(0,a,0); M$_5^+$(0,0;a,0;0,0) \\
(0,a;0,0;0,0$\mid$0;b;0$\mid$0,c,0) & $\Gamma_3^+$(a,b); $\Gamma_4^-$(0,a,0); $\Gamma_5^-$(0,0,a); M$_1^+$(0;a;0); M$_2^+$(0;a;0)\\
(a,$\bar{a}$;0,0;0,0$\mid$0;b;0$\mid$c,c,0) & $\Gamma_3^+$(a,$\sqrt{3}\bar{a}$); $\Gamma_5^+$(0,0,a); $\Gamma_4^-$(0,a,0); M$_1^+$(0;a;0); M$_4^+$(0;a;0); M$_2^-$(0;a;0)\\
\hline
(mX$_1^-$$\mid$M$_5^-$$\mid$mR$_5^-$) & - \\
(a;0;0$\mid$0,0;0,b;0,0$\mid$c,0,0) & $\Gamma_3^+$(a,b); $\Gamma_4^-$(0,a,0); $\Gamma_5^-$(0,0,a); M$_5^+$(0,0;a,0;0,0) \\
(a;0;0$\mid$0,0;0,b;0,0$\mid$0,0,c) & $\Gamma_3^+$(a,b); $\Gamma_4^-$(0,0,a); $\Gamma_5^-$(a,0,0); M$_3^+$(0;a;0); M$_4^+$(0;a;0)\\
(a;0;0$\mid$0,0;b,b;0,0$\mid$0,0,c) & $\Gamma_3^+$(a,$\sqrt{3}\bar{a}$); $\Gamma_5^+$(0,0,a); $\Gamma_4^-$(a,0,a); $\Gamma_5^-$(a,$\bar{a}$,0); M$_1^+$(0;a;0); M$_4^+$(0;a;0)\\
(a;0;0$\mid$0,0;b,b;0,0$\mid$c,c,0) & $\Gamma_3^+$(a,$\sqrt{3}\bar{a}$); $\Gamma_5^+$(0,0,a); $\Gamma_4^-$(0,a,0); M$_5^+$(0,0;a,$\bar{a}$;0,0)\\
\hline
(mX$_5^-$$\mid$M$_5^-$$\mid$mR$_5^-$) & - \\
(0,a;0,0;0,0$\mid$0,0;0,b;0,0$\mid$c,0,0) & $\Gamma_3^+$(a,b); $\Gamma_4^-$(0,0,a); $\Gamma_5^-$(a,0,0); M$_3^+$(0;a;0); M$_4^+$(0;a;0)\\
(0,a;0,0;0,0$\mid$0,0;0,b;0,0$\mid$0,0,c) & $\Gamma_3^+$(a,b); $\Gamma_4^-$(0,a,0); $\Gamma_5^-$(0,0,a); M$_5^+$(0,0;a,0;0,0)\\
(0,a;0,0;0,0$\mid$0,0;0,b;0,0$\mid$0,c,0) & $\Gamma_3^+$(a,b); $\Gamma_4^-$(a,0,0); $\Gamma_5^-$(0,a,0); M$_1^+$(0;a;0); M$_2^+$(0;a;0)\\
(0,a;0,0;0,0$\mid$0,0;$\bar{b}$,0;0,0$\mid$c,0,0) & $\Gamma_3^+$(a,b); $\Gamma_4^-$(a,0,0); $\Gamma_5^-$(0,a,0); M$_3^+$(0;a;0); M$_4^+$(0;a;0)\\
(0,a;0,0;0,0$\mid$0,0;$\bar{b}$,0;0,0$\mid$0,c,0) & $\Gamma_3^+$(a,b); $\Gamma_4^-$(0,0,a); $\Gamma_5^-$(a,0,0); M$_1^+$(0;a;0); M$_2^+$(0;a;0)\\
(a,$\bar{a}$;0,0;0,0$\mid$0,0;b,b;0,0$\mid$0,0,c) & $\Gamma_3^+$(a,$\sqrt{3}\bar{a}$); $\Gamma_5^+$(0,0,a); $\Gamma_4^-$(0,a,0); M$_5^+$(0,0;a,$\bar{a}$;0,0)\\
(a,$\bar{a}$;0,0;0,0$\mid$0,0;b,b;0,0$\mid$c,c,0) & $\Gamma_3^+$(a,$\sqrt{3}\bar{a}$); $\Gamma_5^+$(0,0,a); $\Gamma_4^-$(a,0,a); $\Gamma_5^-$(a,$\bar{a}$,0); M$_1^+$(0;a;0); M$_4^+$(0;a;0)\\
(a,$\bar{a}$;0,0;0,0$\mid$0,0;b,b;0,0$\mid$$\bar{c}$,c,0) & $\Gamma_3^+$(a,$\sqrt{3}\bar{a}$); $\Gamma_5^+$(0,0,a); $\Gamma_4^-$(a,0,$\bar{a}$); $\Gamma_5^-$(a,a,0); M$_2^+$(0;a;0); M$_3^+$(0;a;0)\\
(a,$\bar{a}$;0,0;0,0$\mid$0,0;$\bar{b}$,b;0,0$\mid$c,c,0) & $\Gamma_3^+$(a,$\sqrt{3}\bar{a}$); $\Gamma_5^+$(0,0,a); $\Gamma_4^-$(a,0,$\bar{a}$); $\Gamma_5^-$(a,a,0); M$_1^+$(0;a;0); M$_4^+$(0;a;0)\\
(a,$\bar{a}$;0,0;0,0$\mid$0,0;$\bar{b}$,b;0,0$\mid$$\bar{c}$,c,0) & $\Gamma_3^+$(a,$\sqrt{3}\bar{a}$); $\Gamma_5^+$(0,0,a); $\Gamma_4^-$(a,0,a); $\Gamma_5^-$(a,$\bar{a}$,0); M$_2^+$(0;a;0);M$_3^+$(0;a;0)\\
\hline\hline
\end{tabular}
\end{table}

\begin{table}
\label{SOPsR}
\caption{Structural SOPs of POPs indicated in the table. Polarisation,$\Gamma_{4}^{-}$, is always a SOP.}
\begin{tabular}{lc} % Alignment for each cell: l=left, c=center, r=right
POP & SOPs \\
\hline\hline
(mX$_1^-$$\mid$mM$_2^+$$\mid$R$_5^+$) & - \\
(a;0;0$\mid$0;b;0$\mid$c,0,0) & $\Gamma_3^+$(a,b); $\Gamma_4^-$(0,0,a); $\Gamma_5^-$(a,0,0);R$_2^-$(a); R$_3^-$(a,b)\\
(a;0;0$\mid$0;b;0$\mid$0,0,c) & $\Gamma_3^+$(a,$\sqrt{3}\bar{a}$); $\Gamma_4^-$(0,a,0); R$_2^-$(a); R$_3^-$(a,$\sqrt{3}\bar{a}$)\\
(a;0;0$\mid$0;b;0$\mid$c,c,0) & $\Gamma_3^+$(a,$\sqrt{3}\bar{a}$); $\Gamma_5^+$(0,0,a); $\Gamma_4^-$(a,0,a); $\Gamma_5^-$(a,$\bar{a}$,0); R$_2^-$(a); R$_3^-$(a,$\sqrt{3}\bar{a}$); R$_4^-$(0,0,a)\\
\hline
(mX$_1^-$$\mid$mM$_5^+$$\mid$R$_5^+$) & - \\
(a;0;0$\mid$0,0;b,0;0,0$\mid$0,0,c) & $\Gamma_3^+$(a,b);$\Gamma_4^-$(a,0,0); $\Gamma_5^-$(0,a,0); R$_4^-$(a,0,0); R$_5^-$(a,0,0)\\
(a;0;0$\mid$0,0;b,0;0,0$\mid$0,c,0) & $\Gamma_3^+$(a,b);$\Gamma_4^-$(0,a,0); $\Gamma_5^-$(0,0,a); R$_4^-$(a,0,0);R$_5^-$(a,0,0)\\
(a;0;0$\mid$0,0;b,$\bar{b}$;0,0$\mid$0,0,c) & $\Gamma_3^+$(a,$\sqrt{3}\bar{a}$); $\Gamma_5^+$(0,0,a); $\Gamma_4^-$(a,0,$\bar{a}$);$\Gamma_5^-$(a,a,0);R$_4^-$(a,$\bar{a}$,0); R$_5^-$(a,a,0)\\
(a;0;0$\mid$0,0;b,$\bar{b}$;0,0$\mid$$\bar{c}$,c,0) & $\Gamma_3^+$(a,$\sqrt{3}\bar{a}$);$\Gamma_5^+$(0,0,a); $\Gamma_4^-$(0,a,0); R$_4^-$(a,$\bar{a}$,0); R$_5^-$(a,a,0)\\
\hline
(mX$_5^-$$\mid$mM$_2^+$$\mid$R$_5^+$) & - \\
(a,$\bar{a}$;0,0;0,0$\mid$0;b;0$\mid$0,0,c) & $\Gamma_3^+$(a,$\sqrt{3}\bar{a}$); $\Gamma_5^+$(0,0,a); $\Gamma_4^-$(a,0,a); $\Gamma_5^-$(a,$\bar{a}$,0); R$_4^-$(a,a,0); R$_5^-$(a,$\bar{a}$,0)\\
(0,a;0,0;0,0$\mid$0;b;0$\mid$c,0,0) & $\Gamma_3^+$(a,b); $\Gamma_4^-$(0,a,0); $\Gamma_5^-$(0,0,a); R$_4^-$(0,a,0); R$_5^-$(0,a,0)\\
(0,a;0,0;0,0$\mid$0;b;0$\mid$0,0,c) & $\Gamma_3^+$(a,b); $\Gamma_4^-$(0,0,a); $\Gamma_5^-$(a,0,0); R$_4^-$(0,a,0); R$_5^-$(0,a,0)\\
(a,$\bar{a}$;0,0;0,0$\mid$0;b;0$\mid$c,c,0) & $\Gamma_3^+$(a,$\sqrt{3}\bar{a}$); $\Gamma_5^+$(0,0,a); $\Gamma_4^-$(0,a,0); R$_4^-$(a,a,0); R$_5^-$(a,$\bar{a}$,0)\\
\hline
(mX$_5^-$$\mid$mM$_5^+$$\mid$R$_5^+$) & - \\
(0,a;0,0;0,0$\mid$0,0;b,0;0,0$\mid$c,0,0) & $\Gamma_3^+$(a,b); $\Gamma_4^-$(a,0,0); $\Gamma_5^-$(0,a,0); R$_4^-$(0,0,a); R$_5^-$(0,0,a)\\
(0,a;0,0;0,0$\mid$0,0;b,0;0,0$\mid$0,c,0) & $\Gamma_3^+$(a,b); $\Gamma_4^-$(0,0,a); $\Gamma_5^-$(a,0,0); R$_4^-$(0,0,a); R$_5^-$(0,0,a)\\
(0,a;0,0;0,0$\mid$0,0;0,$\bar{b}$;0,0$\mid$c,0,0) & $\Gamma_3^+$(a,b); $\Gamma_4^-$(0,0,a); $\Gamma_5^-$(a,0,0); R$_2^-$(a); R$_3^-$(a,b)\\
(0,a;0,0;0,0$\mid$0,0;0,$\bar{b}$;0,0$\mid$0,0,c) & $\Gamma_3^+$(a,b); $\Gamma_4^-$(0,a,0); $\Gamma_5^-$(0,0,a); R$_2^-$(a); R$_3^-$(a,b)\\
(0,a;0,0;0,0$\mid$0,0;0,$\bar{b}$;0,0$\mid$0,c,0) & $\Gamma_3^+$(a,b); $\Gamma_4^-$(a,0,0); $\Gamma_5^-$(0,a,0); R$_2^-$(a); R$_3^-$(a,b)\\
(a,$\bar{a}$;0,0;0,0$\mid$0,0;$\bar{b}$,$\bar{b}$;0,0$\mid$0,0,c) & $\Gamma_3^+$(a,$\sqrt{3}\bar{a}$); $\Gamma_5^+$(0,0,a); $\Gamma_4^-$(0,a,0); R$_2^-$(a); R$_3^-$(a,$\sqrt{3}\bar{a}$); R$_4^-$(0,0,a)\\
(a,$\bar{a}$;0,0;0,0$\mid$0,0;b,$\bar{b}$;0,0$\mid$c,c,0) & $\Gamma_3^+$(a,$\sqrt{3}\bar{a}$); $\Gamma_5^+$(0,0,a); $\Gamma_4^-$(a,0,$\bar{a}$); $\Gamma_5^-$(a,a,0); R$_3^-$(a,0.577a); R$_5^-$(0,0,a)\\
(a,$\bar{a}$;0,0;0,0$\mid$0,0;b,$\bar{b}$;0,0$\mid$$\bar{c}$,c,0) & $\Gamma_3^+$(a,$\sqrt{3}\bar{a}$); $\Gamma_5^+$(0,0,a); $\Gamma_4^-$(a,0,a); $\Gamma_5^-$(a,$\bar{a}$,0); R$_3^-$(a,0.577a); R$_5^-$(0,0,a)\\
(a,$\bar{a}$;0,0;0,0$\mid$0,0;$\bar{b}$,$\bar{b}$;0,0$\mid$c,c,0) & $\Gamma_3^+$(a,$\sqrt{3}\bar{a}$); $\Gamma_5^+$(0,0,a); $\Gamma_4^-$(a,0,a); $\Gamma_5^-$(a,$\bar{a}$,0); R$_2^-$(a); R$_3^-$(a,$\sqrt{3}\bar{a}$); R$_4^-$(0,0,a)\\
(a,$\bar{a}$;0,0;0,0$\mid$0,0;$\bar{b}$,$\bar{b}$;0,0$\mid$$\bar{c}$,c,0) & $\Gamma_3^+$(a,$\sqrt{3}\bar{a}$); $\Gamma_5^+$(0,0,a); $\Gamma_4^-$(a,0,$\bar{a}$); $\Gamma_5^-$(a,a,0); R$_2^-$(a); R$_3^-$(a,$\sqrt{3}\bar{a}$); R$_4^-$(0,0,a)\\
\hline\hline
\end{tabular}
\end{table}

    % Postscript figures can be included with multiple figure blocks

\end{document}